\documentclass[journal]{new-aiaa}
\usepackage[utf8]{inputenc}
\usepackage{multicol}
\usepackage{graphicx}
\usepackage{amsmath}
\usepackage[version=4]{mhchem}
\usepackage{siunitx}
\usepackage{algorithm2e}
\RestyleAlgo{ruled}
\usepackage{longtable,tabularx}
\usepackage{color,soul}
\usepackage{subcaption}
\usepackage[section]{placeins} 
\usepackage{titlesec}
\usepackage{hyperref}
\usepackage{algorithm2e}
\usepackage{fancyhdr}
\usepackage{booktabs}
\usepackage{svg}
\SetKwComment{Comment}{// }{}

\title{Density Estimation for Entry Guidance Problems using Deep Learning}
\author{Jens A. Rataczak\footnote{Graduate Research Assistant, AIAA Student Member.}}
\affil{Department of Aerospace Engineering Sciences,\\University of Colorado Boulder, Boulder, CO 80302, USA}
\author{Davide Amato\footnote{Lecturer in Spacecraft Engineering, AIAA Member.}}
\affil{Department of Aeronautics,\\Imperial College London, SW7 2AZ, London, United Kingdom}
\author{Jay W. McMahon\footnote{Associate Professor, AIAA Associate Fellow.}}
\affil{Department of Aerospace Engineering Sciences,\\University of Colorado Boulder, Boulder, CO 80302, USA}

\date{July 2023}

\begin{document}
\maketitle
\begin{abstract}
This work presents a deep-learning approach to estimate atmospheric density profiles for use in planetary entry guidance problems. A long short-term memory (LSTM) neural network is trained to learn the mapping between measurements available onboard an entry vehicle and the density profile through which it is flying. Measurements include the spherical state representation, Cartesian sensed acceleration components, and a surface-pressure measurement. Training data for the network is initially generated by performing a Monte Carlo analysis of an entry mission at Mars using the fully numerical predictor-corrector guidance (FNPEG) algorithm that utilizes an exponential density model, while the truth density profiles are sampled from MarsGRAM. A curriculum learning procedure is developed to refine the LSTM network's predictions for integration within the FNPEG algorithm. The trained LSTM is capable of both predicting the density profile through which the vehicle will fly and reconstructing the density profile through which it has already flown. The performance of the FNPEG algorithm is assessed for three different density estimation techniques: an exponential model, an exponential model augmented with a first-order fading-memory filter, and the LSTM network. Results demonstrate that using the LSTM model results in superior terminal accuracy compared to the other two techniques when considering both noisy and noiseless measurements. 
\end{abstract}

\section*{Nomenclature}

{\renewcommand\arraystretch{1.0}
\noindent\begin{longtable*}{@{}l @{\quad=\quad} l@{}}
$A_{\rm{ref}}$ & reference area, m$^2$ \\ 
$\boldsymbol{a}$ & aerodynamic acceleration vector, m/s\textsuperscript{2} \\ 
$a_x, a_y, a_z$ & aerodynamic acceleration components, m/s\textsuperscript{2} \\ 
$\boldsymbol{b}$ & LSTM bias vector \\
$C_L$ & lift coefficient \\
$C_D$ & drag coefficient \\ 
$\boldsymbol{c}$ & output of LSTM cell state\\ 
$c_0$, $c_1$ & deadband constants, degrees \\ 
$\boldsymbol{D}$ & drag force vector, N \\ 
$D$ & drag force magnitude, N \\ 
$e$ & energy \\ 
$\boldsymbol{F}$ & unknown mapping \\ 
$\boldsymbol{f}$ & output of forget gate \\
$g$ & gravitational acceleration \\ 
$g_0$ & gravitational acceleration at the planet surface \\
$\boldsymbol{h}$ & output of LSTM hidden state \\ 
$h$ & altitude, m \\ 
$\boldsymbol{i}$ & output of input gate \\ 
$K$ & number of trajectories in data set \\ 
$\boldsymbol{L}$ & lift force vector, N \\ 
$L$ & lift force magnitude, N \\ 
$L/D$ & lift-to-drag ratio \\ 
$\mathcal{L}$ & network loss \\ 
$N$ & length of feature vector sequence \\ 
$\boldsymbol{o}$ & output of output gate \\ 
$P$ & pressure, Pa \\ 
$R$ & planet radius \\
$\boldsymbol{r}$ & radius vector, random vector in dropout \\ 
$s$ & range-to-go, deg \\ 
$U$ & LSTM weight matrix \\ 
$\boldsymbol{V}$ & velocity vector, m/s \\ 
$V$ & velocity vector magnitude, m/s \\ 
$W$ & LSTM weight matrix \\ 
$\boldsymbol{x}$ & feature vector \\ 
$z$ & terminal range-to-go error, deg \\ 

\multicolumn{2}{@{}l}{\textit{Greek}}\\
$\beta$ & ballistic coefficient, filter parameter, kg/m$^2$ or ($\cdot$)  \\ 
$\epsilon$ & convergence tolerance \\ 
$\eta$ & pseudodensity \\ 
$\gamma$ & flight-path angle, degrees \\ 
$\lambda$ & step size\\ 
$\mu$ & gravitational parameter of Mars, $4.305\times10^{13}$ m$^3$/s$^2$ \\
$\mu$ & mean \\ 
$\boldsymbol{\Omega}$ & rotation rate vector, deg/s\\
$\Omega$ & rotation rate magnitude of Mars, $4.06\times10^{-3}$ deg/s \\ 
$\phi$ & longitude, degrees \\ 
$\Psi$ & azimuth, degrees \\ 
$\Delta\Psi$ & heading offset, degrees \\
$\psi$ & heading angle, degrees \\ 
$\rho$ & density, kg/m$^3$\\ 
$\rho_L$, $\rho_D$ & aerodynamic force ratio\\
$\Delta\rho$ & density prediction error, kg/m$^3$ \\ 
$\sigma$ & bank angle, sigmoid function, standard deviation \\ 
$\theta$ & latitude, degrees \\ 
$\boldsymbol{\Xi}$ & feature vector sequence \\ 

\multicolumn{2}{@{}l}{\textit{Subscripts}}\\
$0$ & initial/nominal/total \\
$1$,$2$ & pre shock, post shock \\
$c$ & cell state \\ 
$f$ & final, forget gate \\
$i$ & index in feature sequence, input gate\\
$k$ & time step \\ 
$o$ & output gate \\ 
$\infty$ & free stream \\

\multicolumn{2}{@{}l}{\textit{Superscripts}}\\
$j$ & element of feature vector \\ 
$k$ & predictor-corrector iteration, trajectory in data set \\ 
$*$ & guidance target \\
$\dot{}$ & time derivative \\ 
$\tilde{}$ & normalized; filtered \\ 
$\bar{}$ & mean \\
$\check{}$ & standard deviation \\ 

\end{longtable*}}

\section{Introduction}
\lettrine{I}{n} atmospheric entry scenarios, one of the largest sources of uncertainty comes from day-of-flight variation in the atmospheric density profile. For planetary destinations with highly dynamic atmospheric profiles, such as Mars, the density can vary by several orders of magnitude, depending on a variety of factors such as altitude, time of year, proximity to the pole, etc., and such variability in atmospheric density significantly increases mission risk \cite{engelund2008atmospheric}. It is the role of the onboard guidance system to adapt to dispersions in the expected density profile in order to safely guide the entry vehicle to its desired target. State-of-the-art Mars entry descent and landing (EDL) guidance performance was demonstrated during the Mars 2020 mission \cite{way2022assessment, dutta2023postflight}, which landed several kilometers away from its target location. Future crewed missions to Mars will require significantly greater landing accuracy (on the order of \SI{50}{\meter}), to be achieved by vehicles with much larger entry masses ($\sim \SI{40}{\tonne}$). Decreasing the uncertainty in the vehicle's state at the end of the entry phase has the potential to significantly increase the final landing accuracy, as the terminal descent guidance system would not have to adapt to as large of dispersions. This requires the entry guidance to be sufficiently robust to uncertainty in the atmospheric density.

For guided entry systems, such as the recent Mars Science Laboratory \cite{prakash2008mars} and Mars 2020 landers \cite{nelessen2019mars}, the performance of the onboard guidance system is heavily dependent upon the ability to perform accurate \textit{in situ} density estimation measurements. While not yet adopted for Mars missions, most state-of-the-art guidance systems adopt a predictor-corrector architecture \cite{lugo2020overview} in which the onboard computer predicts the remainder of the vehicle's trajectory given the current vehicle state and control profile, computes some error metric in the predicted trajectory, and determines the required changes to the control profile that minimize such error metric. Accurate estimation of the atmospheric density profile greatly improves the prediction phase of these algorithms, as the vehicle dynamics are highly dependent on the density.

The simplest method of modeling the density profile for use in the guidance system is to assume that the density varies exponentially with altitude, where the rate of change of density with altitude is determined by a scale height \cite{Vinh}. Higher fidelity can be achieved by replacing a single exponential density law with a piecewise-continuous exponential, with each exponential segment having its own scale height.
Methods to improve density-profile predictions within the assumption of a single exponential density model have been the focus of numerous previous studies. Perot and Rousseau \cite{perot2002importance} proposed a method that attempts to fit the scale height of the exponential density profile using a least-squares fit of acceleration data available from onboard sensors. Masciarelli et al.~\cite{masciarelli2000analytic} proposed a similar approach that focused on estimating exponential fit parameters using sensed acceleration data and a density scale factor. The use of a scaling factor determined using sensed acceleration data, paired with some sort of smoothing filter, has been a common approach to adapting density estimations in the prediction phase of the guidance algorithm \cite{masciarelli2000analytic, Brunner_2008, Lu_2015}. However, this approach applies a constant correction factor to the expected accelerations, which implicitly assumes that the bias in the truth density profile is constant during the entire prediction phase of the predictor-corrector guidance algorithm. This assumption, particularly if the underlying density model is a single exponential law, can produce significant errors, especially at high altitudes \cite{dutta2014uncertainty, heidrich2022optimal}.
In essence, the main issue of exponential density models is that they do not account for seasonal or geographic variations in density as they do not incorporate any explicit dependence on time, latitude, and longitude.

Recently, more sophisticated approaches to predicting the density in EDL missions have been proposed. Roelke et al.~\cite{roelke2023atmospheric} introduced a filter-based approach that selects a profile from an onboard database based on its correlation to previous density estimations. Tracy et al.~\cite{tracy2023robust} present an additional filter-based method that utilizes a square-root Kalman filter to estimate the ratio between the observed and expected density profiles. Each of these methods is essentially solving an inverse estimation problem, in which measurements available onboard are utilized to estimate the true density profile. Other techniques, such as machine learning, have also proven themselves to be an effective means of solving these types of problems \cite{arridge2019solving}. Wagner et al.~\cite{wagner2011adaptive} generated a machine-learning based approach to forecast Martian atmospheric properties, and Amato and McMahon \cite{Amato_2021} utilized deep learning to reconstruct Martian density and wind profiles. These approaches, while reconstructing seasonal, geographic, and climatic density variations, were not designed to be implemented within a guidance loop for real-time density prediction.

This work seeks to generate a robust and accurate method of predicting density in-the-loop of a predictor-corrector guidance algorithm. A long short-term memory (LSTM) neural network, which is particularly efficient in predicting sequences of data, is trained to learn the mapping between measurements available to a planetary entry vehicle and the density profile through which it is flying. The performance of the guidance algorithm, quantified by the terminal range-to-go, is assessed for the LSTM-based density estimator and existing density estimation techniques using both deterministic and noisy measurements. In doing so, the ability of the machine-learning based approach to improve the accuracy of the atmospheric density estimate within an entry guidance algorithm is demonstrated. The increased accuracy, in turn, has the potential to improve the robustness of EDL for future planetary exploration missions.

The paper is organized as follows. Section \ref{sec:lstm} presents the fundamental aspects of deep learning for regression problems and of the specific LSTM architecture used. The physical models, including equations of motion and guidance algorithm, are discussed in Section \ref{sec:physics}. An overview of the training data generation, LSTM training, and of its integration in the FNPEG guidance loop through a curriculum learning procedure are provided in Section \ref{sec:train}. A comparison of the guidance performance using the LSTM-based density estimator to existing density estimation techniques, with and without noise, is presented in Section \ref{sec:results}, and conclusions are presented in Section \ref{sec:conclusions}.

\section{Neural Network Architecture}\label{sec:lstm}
This section introduces neural networks concepts that are central to this work. The general non-linear regression problem is reviewed, and the pitfalls of various neural networks for time-varying data are examined. A general overview of the LSTM neural network is provided in \ref{subsec:lstm}, and the concept of dropout layers for improved generalization is described in \ref{subsec:dropout}.
\subsection{Long Short-Term Memory Neural Network}\label{subsec:lstm}
\begin{figure}[htbp]
    \centering
    \includegraphics[width=0.7\textwidth]{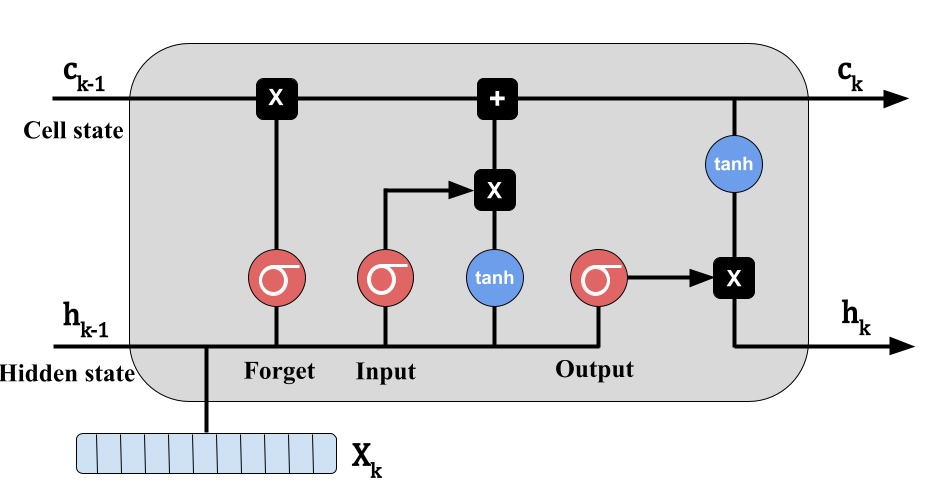}
    \caption{Schematic of LSTM cell.}
    \label{fig:lstm_cell}
\end{figure}

For a general regression problem, a neural network can be trained to model some unknown, nonlinear mapping $\boldsymbol{F}(\boldsymbol{x}): \mathbb{R}^n\mapsto\mathbb{R}^m$. In the context of density estimation for entry guidance, the mapping $\boldsymbol{F}$ represents the nonlinear relationship between measurements taken by a spacecraft and the corresponding density profile through which it is flying. Classical deep feed-forward neural networks (FFNN) are highly capable of approximating nonlinear functions. However, they are not as efficient at making predictions from \textit{sequences} of measurements of arbitrary length, as an FFNN would need to be re-trained if the length of the input sequence varies. This issue can be avoided by employing recurrent neural networks (RNNs), which use the same weights across time steps and do not need to be re-trained for different sequence lengths and are thus more efficient at predicting targets from sequences of data of arbitrary length \cite{Goodfellow-et-al-2016}. This capability is particularly attractive because the length of the measurement sequence used to estimate the density grows continuously along the trajectory. However, traditional RNNs, in which the input at each time step $\boldsymbol{x}_k$ is passed through a simple FFNN, are very hard to train on long sequences of data. This is due to the ``vanishing gradient'' phenomenon: the gradient of the loss function rapidly decreases as it is propagated backwards along the time steps of the sequence, resulting in a loss of prediction capability for increasingly long sequences \cite{Hochreiter_1998}. One way of mitigating the vanishing gradient is to replace the standard hidden layers in a classical RNN with a deep subnetwork called the long short-term memory (LSTM) cell \cite{Hochreiter_1997}. The LSTM cell aids in the learning of long-term dependencies by introducing three hidden layers, denominated as ``gates''. Specifically, these are the ``input'', ``forget'', and ``output'' gates which control the flow of information through the cell. The input gate determines which information will be stored in the cell, the forget gate determines which information will be removed or ``forgotten'' from the cell, and the output gate determines what information will be output from the LSTM cell. A generic schematic of the LSTM cell is shown in Figure \ref{fig:lstm_cell}.

At a given time step $k$ in the sequence of features with corresponding network input $\boldsymbol{x}_k$, the outputs $\boldsymbol{f}_k$, $\boldsymbol{i}_k$, $\boldsymbol{o}_k$, of the forget, input, and output gates, respectively, are governed by
\begin{eqnarray}
    \boldsymbol{f}_k&=& \sigma\left(W_f\cdot\boldsymbol{x}_k+ U_f\cdot\boldsymbol{h}_{k-1} + \boldsymbol{b}_f\right)  \\
    \boldsymbol{i}_k&=& \sigma\left(W_i\cdot\boldsymbol{x}_k+ U_i\cdot\boldsymbol{h}_{k-1} + \boldsymbol{b}_i\right) \\
    \boldsymbol{o}_k&=& \sigma\left(W_o\cdot\boldsymbol{x}_k+ U_o\cdot\boldsymbol{h}_{k-1} + \boldsymbol{b}_o\right) \qquad, 
\end{eqnarray}
 where the matrices $W_{(\cdot)}\in\mathbb{R}^{h\times d}$, $U_{(\cdot)}\in\mathbb{R}^{h\times h}$ and vectors $\boldsymbol{b}_{(\cdot)}\in\mathbb{R}^h$ are weight matrices and bias vectors, respectively, to be learned during training. The vector $\boldsymbol{h}_{k-1}\in\mathbb{R}^h$ is the hidden state of the LSTM cell at the previous time step. The mapping $\sigma(x)\mapsto \dfrac{1}{1+e^{-x}}$ is called the ``sigmoid'' function and is a common transformation applied to the output of a neural network layer, as it maps the real line to the interval (0,1). The outputs $\boldsymbol{c}_k$ and $\boldsymbol{h}_k$ of the cell state and hidden state (output of LSTM unit), respectively, of the LSTM are then given by 
\begin{eqnarray}
    \boldsymbol{c}_k &=& \boldsymbol{f}_k \odot \boldsymbol{c}_{k-1} + \boldsymbol{i}_k\odot \tanh\left(W_c\cdot\boldsymbol{x}_k + U_c\cdot\boldsymbol{h}_{k-1} + \boldsymbol{b}_c\right)  \\ 
    \boldsymbol{h}_k &=& \boldsymbol{o}_k\odot\sigma(\boldsymbol{c}_k)\qquad ,
\end{eqnarray}
where, $W_c\in\mathbb{R}^{h\times d}$, $U_c\in\mathbb{R}^{h\times h}$, and $\boldsymbol{b}_c\in\mathbb{R}^{h}$ are additional parameters to be learned during training, and the operation $\odot$ represents element-wise multiplication. For a more in-depth treatment of the LSTM algorithm, the reader is referred to Reference \cite{Sherstinsky_2020}.

Extending upon the work in \cite{Amato_2021}, this work implements a long short-term memory (LSTM) neural network to learn the relationship between an entry vehicle's dynamics and the density profile through which it is flying. One of the primary differences between this work and Reference \cite{Amato_2021} is that the LSTM in this work performs ``sequence-to-sequence'' mapping, where the LSTM outputs a vector sequence of the same length as the sequence of input feature vectors (i.e. for all time steps leading up to the current time step) \cite{Sutskever_2014}, whereas the network architecture in \cite{Amato_2021} was for ``sequence-to-one'' mapping in which the LSTM only produces an output corresponding to the current time step. A sequence-to-sequence architecture is chosen because accurate density predictions are desired at every time step, not just the at the final time step, for the guidance algorithm to function optimally.

\subsection{Dropout Layers}\label{subsec:dropout}
\begin{figure}[htbp]
    \centering
    \includegraphics[width=0.7\textwidth]{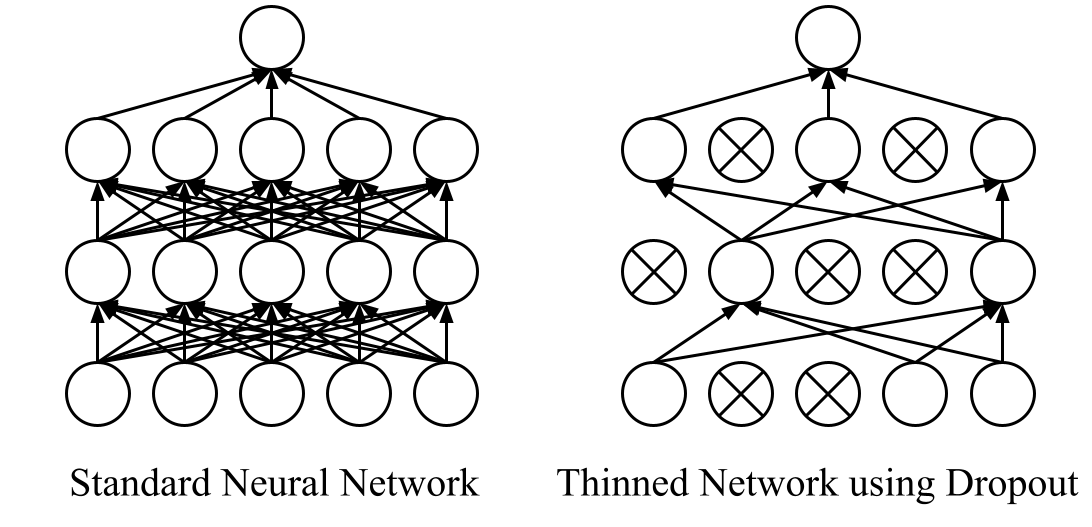}
    \caption{Standard neural network (left) and a thinned neural network after applying dropout (right).}
    \label{fig:dropout}
\end{figure}

Deep neural networks are powerful tools for function approximation; however, a common concern is the concept of ``overfitting.'' A network is said to be overfit, or overtrained, if it only produces accurate predictions for the data set on which it was trained, meaning that it does not have the ability to generalize its predictive capabilities to data that was not used to train the model. In the context of this work, this means that the network would fail to accurately predict density profiles that were not included in the training set. A common technique to avoid overfitting and improve the network's ability to generalize to new data is to introduce ``dropout'' layers \cite{Goodfellow-et-al-2016}. Dropout is a stochastic regularization technique that is equivalent to adding noise to the output of the hidden units of the network \cite{Srivastava_2014}. A dropout layer randomly drops (i.e. sets to zero) some of the previous layer's outputs (see Figure \ref{fig:dropout}) to prevent the network from relying too heavily on the activation of specific connections within the network and focus more on probabilistically satisfying the unknown non-linear mapping. This signifies that dropout helps reduce the likelihood of layers within the network ``working together'' to correct mistakes from prior layers, in turn making the model more robust.

 Dropout is achieved by sampling a random vector from a Bernoulli distribution and multiplying it element-wise with the output of a network layer. The LSTM output $\boldsymbol{h}_k$ is modified via
\begin{eqnarray}
    \tilde{\boldsymbol{h}}_k &=& \boldsymbol{r}\odot\boldsymbol{h}_k\qquad,
\end{eqnarray}
where $\boldsymbol{r} \in \mathbb{R}^h$ is a random vector sampled from a Bernoulli distribution with probability $p$. It is important to note that using dropout layers only affects the network structure during training. When the network is performing inference (during validation/testing), the dropout layers are not active.

\begin{figure}[htbp]
    \centering
    \includegraphics[width=\textwidth]{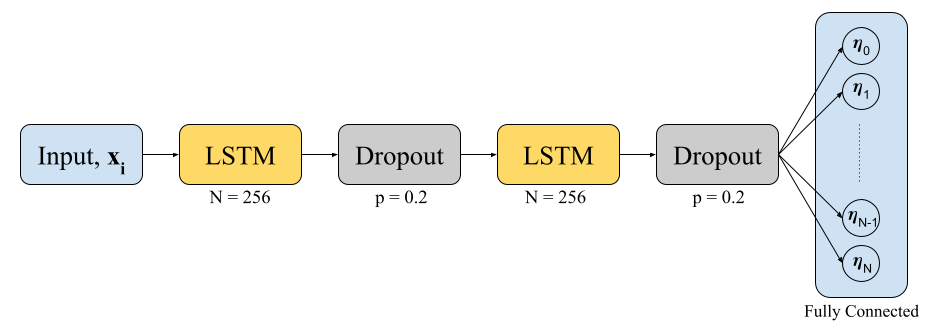}
    \caption{LSTM network architecture used in this work.}
    \label{fig:lstm}
\end{figure}

The network architecture used in the rest of this work is shown in Figure \ref{fig:lstm}. Two LSTM layers, with 256 hidden units each, both followed by a dropout layer with a drop rate of $p=0.2$, comprise the hidden layers of the network, and the output of the second dropout layer is passed through a fully-connected layer to transform it into the sequence of target vectors. The choice to use two LSTM layers comes from the fact that the network is trying to predict perturbed density profiles, so one LSTM layer can identify the mean density profile and a second LSTM layer can identify the perturbations about that mean profile. However, this was not enforced explicitly within the training.

\section{Physical Models}\label{sec:physics}
The equations of motion governing the dynamics of a Martian entry vehicle are presented in \ref{subsec:dynamics}, and the guidance algorithm used to guide the vehicle from its entry state to a target location is described in \ref{subsec:fnpeg}.

\subsection{Dynamics}\label{subsec:dynamics}
The motion of the entry vehicle is expressed using Cartesian equations of motion defined in a Mars-centered, Mars-fixed reference frame with its first axis pointing toward the intersection of the prime meridian with the equator, the third axis aligned with Mars' axis of rotation, and the second axis completing the right-hand set. The equations of motion are
\begin{subequations}\label{eq:eoms_cart}
    \begin{eqnarray}
    \dot{\boldsymbol{r}} &=& \boldsymbol{V} \label{eq:rdot_cart}\\
    \dot{\boldsymbol{V}} &=& -\frac{\mu}{r^3}\boldsymbol{r} + \boldsymbol{L} + \boldsymbol{D} - 2\boldsymbol{\Omega}\times\boldsymbol{V} - \boldsymbol{\Omega}\times\left(\boldsymbol{\Omega}\times\boldsymbol{r}\right)\qquad ,\label{eq:vdot_cart}
\end{eqnarray}
\end{subequations}
where $\boldsymbol{r}$ is the radius vector, $\boldsymbol{V}$ is the planet-relative velocity vector, $\mu$ is the gravitational constant for Mars, and the angular velocity vector $\boldsymbol{\Omega}=\left(0,~0,~\Omega\right)$. The lift and drag acceleration vectors act normal and antiparallel, respectively, to the velocity vector:
\begin{eqnarray}
    \boldsymbol{L} &=& L\left( \cos\sigma\frac{\boldsymbol{V}\times\left(\boldsymbol{r}\times\boldsymbol{V}\right)}{V||\boldsymbol{r}\times\boldsymbol{V}||} + \sin\sigma\frac{\boldsymbol{r}\times\boldsymbol{V}}{||\boldsymbol{r}\times\boldsymbol{V}||} \right), \label{eq:lift_cart}\\
    \boldsymbol{D} &=& -D\frac{\boldsymbol{V}}{V}\qquad,
\end{eqnarray}
where the bank angle $\sigma$ is the angle between the local vertical plane containing the velocity vector and the plane containing the velocity and the aerodynamic force vectors (i.e. angle of rotation about the velocity vector) \cite{Vinh}. 
The lift and drag acceleration magnitudes are determined with Equations \ref{eq:lift} and \ref{eq:drag}, respectively:
\begin{eqnarray}
    L &=& \frac{1}{2}\rho_\infty V^2 C_L A_{\rm{ref}} = \frac{\rho_\infty V^2}{2\beta}  \label{eq:lift}\\
    D &=& \frac{1}{2}\rho_\infty V^2 C_D A_{\rm{ref}} =  \frac{\rho_\infty V^2 L/D}{2\beta} \qquad . \label{eq:drag}
\end{eqnarray}
Here, $\rho_\infty$ is the local freestream density, $L/D = \dfrac{C_L}{C_D}$ is the lift-to-drag ratio, and $\beta = \dfrac{m}{C_D A_{\rm{ref}}}$ is the ballistic coefficient. To include representative variations in the Martian atmosphere, the 2021 Mars Global Reference Atmospheric Model (MarsGRAM) \cite{mars_gram2021} is used to compute the atmospheric properties as a function of altitude, latitude, and longitude.

The entry vehicle is modeled as a hypersonic inflatable aerodynamic decelerator (HIAD) \cite{hughes2011hypersonic}. The physical properties of the HIAD are summarized in Table \ref{tab:hiad}. The ballistic coefficient and $L/D$ values were held constant throughout the simulation, as the vehicle spends most of its time in the hypersonic regime (Mach $\gg 1$) where a constant aerodynamic coefficient assumption is valid \cite{Anderson_2019_ch3}, and the vehicle is not losing any mass.
\begin{table}[htbp]
    \centering
    \caption{HIAD physical properties.}
    \begin{tabular}{l c}
    \toprule\toprule
         Entry Mass (kg) &  $4.9\times10^4$ \\
         Ballistic Coefficient (kg/m$^2$) & 155 \\
         Trim $L/D$ & 0.15 \\ \bottomrule \bottomrule
    \end{tabular}
    \label{tab:hiad}
\end{table}

\subsection{Fully-Numerical Predictor-Corrector Entry Guidance Algorithm}\label{subsec:fnpeg}
Introduced by Lu \cite{Lu_2014}, the fully-numerical predictor-corrector entry guidance (FNPEG) algorithm is a state-of-the-art bank-angle entry guidance algorithm. FNPEG expresses the dynamics of the vehicle in spherical coordinates
\begin{subequations}\label{eq:sph_dynamics}
\begin{align}
\Dot{r} &= V\sin\gamma \qquad \label{eq:rdot} \\
\Dot{\theta} &= \frac{V\cos\gamma\sin\psi}{r\cos\phi} \qquad  \\
\Dot{\phi} &= \frac{V\cos\gamma\cos\psi}{r} \qquad \\
\Dot{V} &= -D - g\sin\gamma+\Omega^2r\cos\phi\left(\sin\gamma\cos\phi - \cos\gamma\sin\phi\cos\psi\right) \qquad  \label{eq:vdot}\\
V\Dot{\gamma} &= L\cos\sigma - g\cos\gamma + \frac{V^2}{r}\cos\gamma + 2\Omega V\cos\phi\sin\psi+ \Omega^2 r\cos\phi\left(\cos\gamma\cos\phi+\sin\gamma\cos\psi\sin\phi\right)  \label{eq:gammadot}\\
V\Dot{\psi} &= \frac{L\sin\sigma}{\cos\gamma} + \frac{V^2}{r}\cos\gamma\sin\psi\tan\phi - 2\Omega V\left(\tan\gamma\cos\psi\cos\phi - \sin\phi\right) + \frac{\Omega^2 r}{\cos\gamma}\sin\psi\sin\phi\cos\phi \quad \label{eq:psidot}
\end{align}
\end{subequations}
Here, $r$ is the vehicle position radius relative to the center of the Mars, $\theta$ is the longitude, $\phi$ is the latitude, $V$ is the planet-relative velocity, $\gamma$ is the flight path angle (defined positive above local horizon), $\psi$ is the heading angle (measured clockwise in the local horizontal plane from the north), and $L$ and $D$ are the lift and drag acceleration magnitudes, respectively. 

Overall, the goal of FNPEG is to determine a bank-angle profile that guides the vehicle from its current state to a target location. The algorithm is split into longitudinal and lateral channels. The longitudinal channel considers only the variables ($r,~V,~\gamma$), described by Equations \ref{eq:rdot}, \ref{eq:vdot}, and \ref{eq:gammadot}. After neglecting the rotation rate $\Omega$ and non-dimensionalizing the longitudinal variables as $\tilde{r}=r/R$, $\tilde{V}=V/\sqrt{g_0R}$, $\tilde{L}=L/g_0$, and $\tilde{D}=D/g_0$, where $R$ is the radius of the planet and $g_0$ is the acceleration due to gravity at the planet surface, the longitudinal equations of motion become
\begin{subequations}\label{eq:lon_dyn}
\begin{eqnarray}
    \Dot{\tilde{r}} &=& \tilde{V}\sin\gamma \qquad \label{eq:rdot_lon} \\
    \Dot{\tilde{V}} &=& -\tilde{D} - \frac{\sin\gamma}{\tilde{r}^2} \qquad  \\
    \tilde{V}\Dot{\gamma} &=& \tilde{L}\cos\sigma - \frac{\cos\gamma}{\tilde{r}^2} + \frac{\tilde{V}^2}{\tilde{r}}\cos\gamma \qquad \label{eq:gammadot_lon} \quad .
\end{eqnarray}
\end{subequations}
Note that the gravitational parameter $\mu$ becomes unity as a consequence of the non-dimensionalization. The magnitude of the bank angle $\sigma$ is prescribed as a linear function of an energy-like variable $e=\dfrac{1}{\tilde{r}}-\dfrac{\tilde{V}^2}{2}$ such that
\begin{equation}
    |\sigma(e)| = \sigma_0 + \frac{e-e_0}{e_f-e_0}\left(\sigma_f-\sigma_0\right)\qquad \quad, \label{eq:bank}
\end{equation}
where $e_0$ and $e_f$ are defined by the current conditions $(\tilde{r},~\tilde{V})$ and target final conditions $(\tilde{r}^*,~\tilde{V}^*)$, respectively. The value $\sigma_f$ is a tuning parameter and is held constant at $\sigma_f=70$ degrees in this work \cite{Lu_2008}. The goal of the longitudinal channel is to determine a value of $\sigma_0\geq0$ that minimizes the error in range-to-go at the final energy with respect to a specified range-to-go,
\begin{eqnarray}
    z(\sigma_0) &=& s(e_f) - s_f^* \quad,
\end{eqnarray}
where the range $s$ is computed by numerically integrating
\begin{equation}
    \dot{s} = -\frac{\tilde{V}\cos\gamma}{\tilde{r}}\label{eq:sdot} \quad.
\end{equation}
The current bank angle command $\sigma_0$ at each call of the guidance algorithm is computed through a step-size controlled Newton-Rhapson method. At the $k$-th iteration of the method, the predictor step numerically integrates Equations \ref{eq:rdot_lon}--\ref{eq:gammadot_lon} and \ref{eq:sdot}, from $e_0$ to $e_f$, using the bank-angle profile defined in Equation \ref{eq:bank} with the current guess $\sigma_0^{(k)}$. In the correction step, a new approximation $\sigma_0^{(k+1)}$ is computed through
\begin{equation}
    \sigma_0^{(k+1)} = \sigma_0^{(k)} - \lambda^{(k)}\frac{z(\sigma_0^{(k)})}{\partial z(\sigma_0^{(0)})/\partial\sigma_0}\quad,
\end{equation}
where the partial derivative is first computed using finite differences for $k=1$ and then switches to a secant method for $k\geq2$. The step size is taken to be $\lambda^{(k)}=1/2^i$ where $i\in\mathbb{Z}_+$ is the smallest value such that $\left|z(\sigma_0^{(k+1)})\right|<\left|z(\sigma_0^{(k)})\right|$. The predictor-corrector process is repeated until the following stopping condition is met:
\begin{equation}
    \left|z(\sigma_0^{(k+1)})\frac{\partial z(\sigma_0^{(k+1)})}{\partial\sigma_0}\right|\leq \epsilon \quad, \label{eq:stop}
\end{equation}
where $\epsilon = 10^{-6}$ in this work.

Note that Equation \ref{eq:sdot} is only valid if the offset between the current heading angle $\psi$ and the azimuth $\Psi$ defining the great circle from the current location to the target is small. To keep this offset small, the lateral guidance channel defines a velocity-dependent deadband such that if 
\begin{equation}
    \left|\Psi - \psi\right| \geq \Delta\Psi = c_1V + c_0 \quad,
\end{equation}
a bank-angle reversal is commanded. The coefficients $c_0$ and $c_1$ are chosen to minimize the number of bank-angle reversals. The FNPEG algorithm is called once every second if the sensed aerodynamic loading $\sqrt{L^2+D^2} \leq 1.47~\rm{m/s^2}$ (0.15 g). If not, the dynamic pressure is too low for the aerodynamic control to be effective. The FNPEG parameters used for all simulations in this work are summarized in Table \ref{tab:FNPEG}.
\begin{table}[htbp]
    \centering
    \caption{Parameters used in FNPEG.}
    \begin{tabular}{l c}
    \toprule\toprule
         Frequency & 1 Hz  \\
         $\sigma_f$ & 70 deg \\
         $\epsilon$ & $10^{-6}$ \\ 
         $\Delta\Psi_0$ & 2.0 deg \\
         $\Delta\Psi_f$ & 1.5 deg \\ \bottomrule\bottomrule
    \end{tabular}
    \label{tab:FNPEG}
\end{table}

\section{LSTM-Augmented FNPEG Guidance}\label{sec:train}
Generally, the process for training a neural network may be broken into three steps. The first entails the generation of the data used to train the network. This step is detailed in Sections \ref{subsec:data_set_gen} and \ref{subsec:data_set_prep}. Next, the network is trained by optimizing the weights and biases of each layer. This is first done by considering an exponential density law in FNPEG in \ref{subsec:train}, and the utility of the network is quantified by testing its predictive capabilities on data in a test set. Finally, the network is trained for implementation into a LSTM-augmented FNPEG guidance algorithm through a curriculum learning approach in \ref{subsec:LSTM_FNPEG}. 

\subsection{Data Set Generation}\label{subsec:data_set_gen}
\begin{figure}[htbp]
\begin{subfigure}{0.49\textwidth}\centering
  \includegraphics[width=\textwidth]{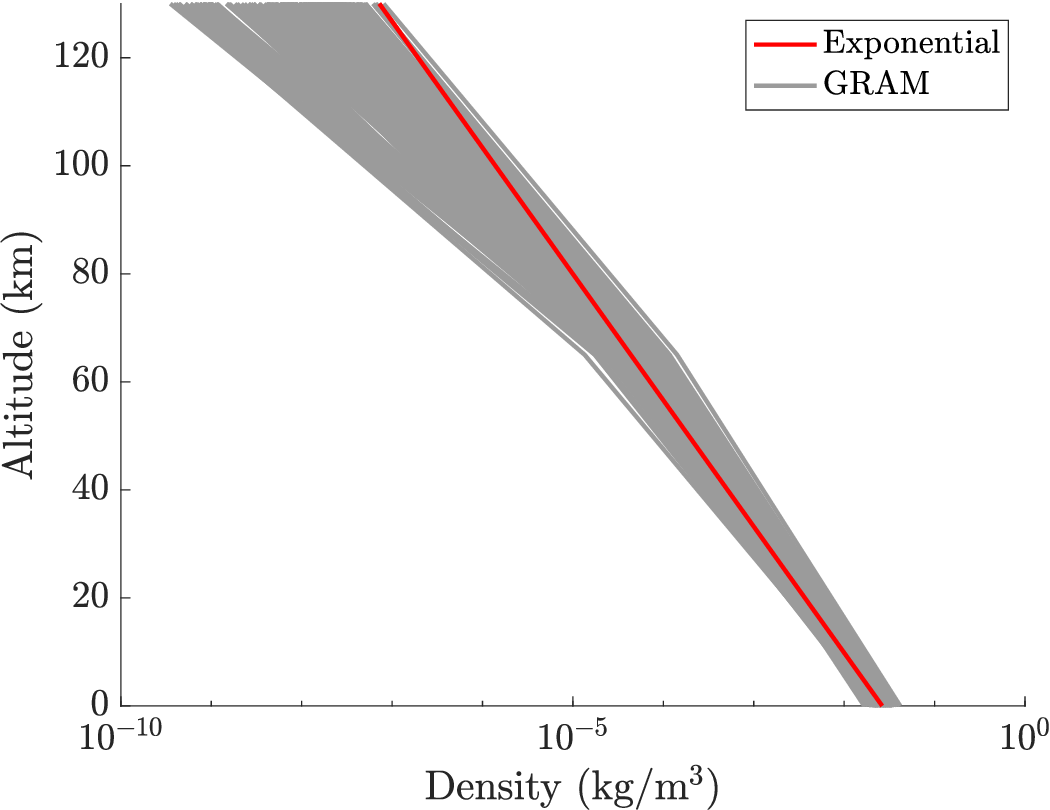}  
  \caption{Density profiles.}
  \label{fig:gram_dens}
\end{subfigure}
\hfill
\begin{subfigure}{0.49\textwidth}\centering
  \includegraphics[width=\textwidth]{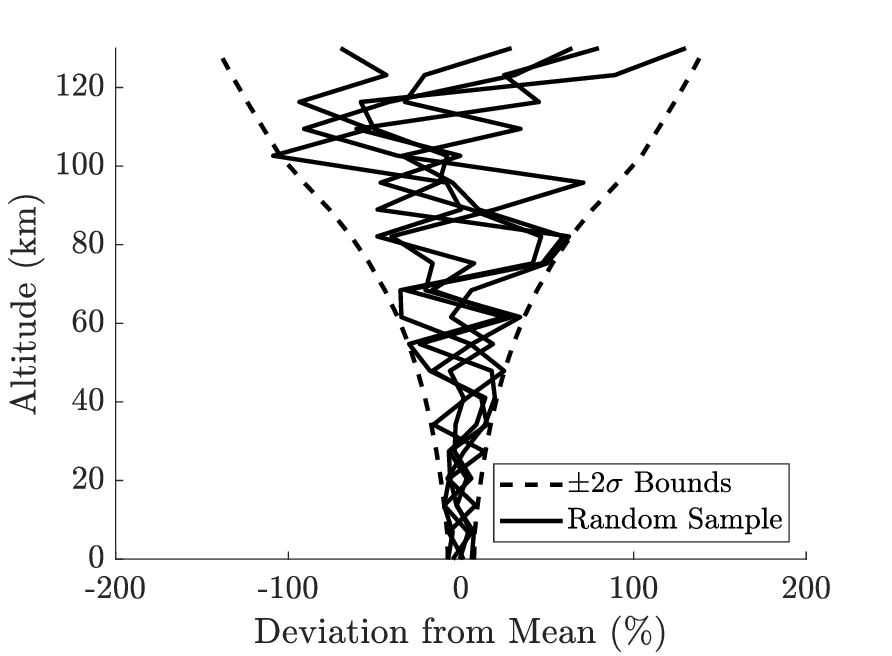}  
  \caption{Deviation from mean.}
  \label{fig:gram_dev}
\end{subfigure}
\caption{Sample perturbed density profiles from MarsGRAM and nominal exponential model (left) and deviation of perturbed density profile from mean (right).}
\label{fig:gram_profs}
\end{figure}

The reference mission (taken from \cite{Amato_2021}), common to all simulations in the training data set, represents the entry phase of a high-mass mission to Mars and is summarized in Table \ref{tab:ref_mission}.
\begin{table}[htbp]
    \centering
    \caption{Initial and target conditions of reference mission.}
    \begin{tabular}{l c c}
    \toprule\toprule
        & Initial & Target \\ \midrule
         Altitude, $h$ (km) & 130 & 11  \\
         Velocity, $V$ (km/s) & 4.0  & 1.214\\
         Longitude, $\theta$ (deg) & 90 & 101.031\\
         Latitude, $\phi$ (deg) & 45 & 47.203 \\
         Range-to-go, $s$ (deg) & 7.952 & 0 \\
         \bottomrule\bottomrule
    \end{tabular}
    \label{tab:ref_mission}
\end{table}
\newpage
To generate the initial training data, 5,000 individual trajectories are simulated. In each simulation, the truth atmospheric profile is generated using MarsGRAM, while the FNPEG algorithm utilizes a simple exponential density model 
\begin{equation}
    \rho(h) = 2.63\times10^{-2}\exp(-h/10.15~\rm{km}) \; \text{kg/m\textsuperscript{3}}\label{eq:exp_dens}\quad.
\end{equation}
Each trajectory is integrated from the initial conditions in Table \ref{tab:ref_mission} until the final energy $e_f$ is reached. Every simulation varied the truth density profile by drawing random samples for the dust level, mean wave offset, and GRAM random seed to simulate variations in the Martian atmosphere due to known physical mechanisms. The distributions from which these samples were drawn are summarized in Table \ref{tab:gram}, which were chosen to produce the largest variation in atmospheric density while still maintaining the feasibility of the guidance problem (i.e. \textit{if} the guidance system had perfect knowledge of the density, the FNPEG algorithm would converge on a bank-angle profile that drives the range-to-go to zero). GRAM is capable of producing density profiles affected by additive random perturbations. The magnitude of the density perturbation is proportional to the density perturbation scale parameter\footnote{This was known as $\tt rpscale$ in older versions of GRAM.} $\delta\rho\in[0,~2]$. Here, all simulations  considered $\delta\rho=2$ to account for maximal variation in the atmospheric profiles. Figure \ref{fig:gram_profs} shows a representative distribution of density profiles sampled from MarsGRAM using the distributions in Table \ref{tab:gram} and the corresponding deviation of the perturbed density profile from the mean density profile. The exponential density model is overlaid for comparison.
\begin{table}[htbp]
    \centering
    \caption{Distributions in MarsGRAM settings.}
    \begin{tabular}{l c}
    \toprule\toprule
        &  Distribution/Value \\ \midrule
         Dust Level & $\mathcal{U}(0.1, 3.0)$  \\
         Mean Wave Offset &  $\mathcal{U}(1.5, 2.5)$\\
         Random Seed & $\mathcal{U}(1, 9\times10^{8})$ \\
         Density Perturbation Scale & 2 \\
         \bottomrule\bottomrule
    \end{tabular}
    \label{tab:gram}
\end{table}

\subsection{Preparation of Data for Training}\label{subsec:data_set_prep}
For each of the 5,000 trajectories in the data set, the true density from MarsGRAM is recorded as a function of time, altitude, latitude, and longitude. However, because the altitude decreases monotonically with time, the density may be expressed solely as a function of altitude using an appropriate change of variables
\begin{equation}
    \rho = \rho(h, t(h), \phi(h), \theta(h))\quad .
\end{equation}
As density varies exponentially with altitude, an additional transformation is introduced
\begin{equation}
    \eta(h) = \sqrt{-\rm{log}_{10}\rho(h)} \quad, \label{eq:pseudo}
\end{equation}
where $\eta$ is called a \textit{pseudodensity}. This transformation produces network targets that vary linearly with altitude and are therefore better conditioned for network training. Note that, at Mars, $-\log_{10}\rho(h) \geq 0~\forall~h$ when $\rho$ is expressed in \text{kg/m\textsuperscript{3}}, so Equation \ref{eq:pseudo} is always well-defined.\footnote{It is possible to keep Equation \ref{eq:pseudo} well-defined for any unit of measurement of density simply through the addition of a constant to the radicand.} For each trajectory, the full profile of pseudodensities $\eta(h)$ is interpolated onto a discrete set of prediction altitudes $\bar{h} = \left\{h_j~|~ h_j = h_{\rm{min}} + \Delta h(j-1),~4\leq h_j \leq 80~\rm{km},~\Delta h = 2~\rm{km} \right\}$. If a trajectory terminates at an altitude $h_f>h_{\rm{min}}$, the density for all $h_j < h_f$ is computed via linear extrapolation.

The input to the network is a sequence of feature vectors $\boldsymbol{\Xi} = \left(\boldsymbol{x}_0,\ldots,\boldsymbol{x}_i\right)$. At the $i$-th time step in the trajectory, the $i$-th vector in the feature vector sequence is defined as
\begin{equation}
    \boldsymbol{x}_i = \left(r_i,~\theta_i, ~\phi_i,~V_i,~\gamma_i,~\psi_i, a_x,~a_y,~a_z,\rm{log}_{10}P_{0,2}\right) \quad.
\end{equation}
Here, the first six elements correspond to the spherical coordinate representation of the vehicle's Cartesian state, the aerodynamic acceleration $\boldsymbol{a} = \boldsymbol{L}+\boldsymbol{D} = \left(a_x,~a_y,~a_z\right) $ is the sensed acceleration measured by the onboard inertial measurement unit (IMU), and the last element of the feature vector is representative of a surface-pressure measurement recorded by a pressure transducer like that found on the Mars Entry Descent and Landing (MEDLI) suite \cite{Hwang_2016, White_2022}. The surface-pressure measurement is approximated by computing the post-shock stagnation pressure using normal-shock relations (Equation \ref{eq:pressure}) and the assumption that pressure is constant across the boundary layer \cite{Anderson_2019_ch3}
\begin{equation}
    P_{0,2} = P_{0,1}\left[ \frac{\left(\gamma+1\right)M_\infty^2}{(\gamma-1)M_\infty^2 + 2} \right]^{\frac{\gamma}{\gamma -1}}\left[\frac{\gamma+1}{2\gamma M_\infty^2-(\gamma-1)}\right]^{\frac{1}{\gamma-1}} \label{eq:pressure} \quad.
\end{equation}
Here $P_{0,1}$ is the freestream stagnation pressure, $M_\infty$ is the freestream Mach number, and $\gamma=1.28$ is the ratio of specific heats for the Martian atmosphere.

To improve training convergence, the feature sequences and target vectors are normalized using z-score normalization (i.e zero mean and unit variance) \cite{Montavon_2012}. For the $k$-th sample in the data set, the $j$-th element of the $i$-th normalized feature vector $^j\Tilde{x}_i^{(k)}$ is computed by
\begin{equation}
    ^j\Tilde{x}_i^{(k)} = \frac{ ^jx^{(k)}_i - {}^j\Bar{x} }{^j\check{x}} \quad,
\end{equation}
where $^j\Bar{x}$ and $^j\check{x}$ are the mean and standard deviation, respectively, of the $j$-th component of the feature vector computed over all time steps and samples
\begin{eqnarray}
    ^j\Bar{x} &=& \frac{1}{K}\sum_{k=1}^K \frac{1}{N(k)}\sum_{i=1}^{N(k)} ~^jx^{(k)}_i  \\
    ^j\check{x} &=& \sqrt{\frac{1}{K}\sum_{k=1}^K \frac{1}{N(k)}\sum_{i=1}^{N(k)}\left(^j x^{(k)}_i- {}^j\Bar{x}\right)^2}\quad.
\end{eqnarray}
Here, $K=\num{5000}$, and $N(k)$ is the number of time steps in $k$-th trajectory in the data set. The target vectors are normalized in a similar manner
\begin{equation}
    ^j\tilde{\eta}^{(k)} = \frac{ ^j\eta^{(k)} - {}^j\Bar{\eta} }{^j\check{\eta}} \quad,
\end{equation}
where the mean and standard deviation are computed over all samples
\begin{eqnarray}
    ^j\Bar{\eta} &=& \frac{1}{K}\sum_{k=1}^K ~^j\eta^{(k)} \\
    ^j\check{\eta} &=& \sqrt{\frac{1}{K}\sum_{k=1}^K \left(^j \eta^{(k)}- {}^j\Bar{\eta}\right)^2}\quad.
\end{eqnarray}
With this, the goal of the network training is to determine the unknown mapping $\Tilde{\boldsymbol{\eta}} = \boldsymbol{F}(\tilde{\boldsymbol{x}})$, which transforms measurement sequences into pseudo-density profiles.

\subsection{Training Results}\label{subsec:train}
\begin{figure}[htbp]
    \centering
    \includegraphics[width=0.6\textwidth]{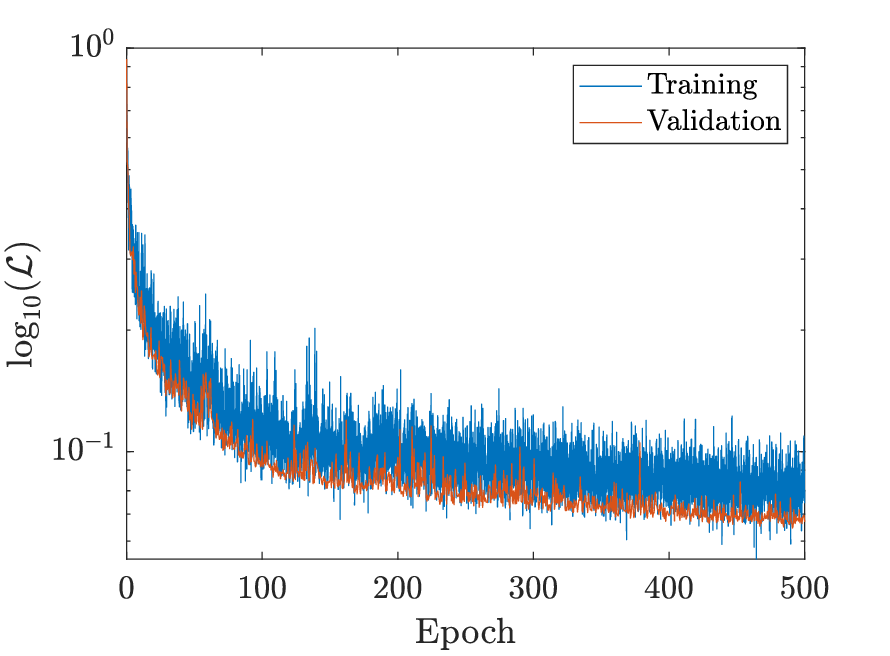}
    \caption{Training and validation loss during 500-epoch training period.}
    \label{fig:train_loss}
\end{figure}

The neural network architecture described in Section \ref{sec:lstm} is trained on 4000 trajectories from the data set in Section \ref{subsec:data_set_prep} using the MATLAB 2022b Deep Learning toolbox. The remaining 1000 trajectories were reserved for the validation data set. Both the training and validation data sets are partitioned into mini batches with a size of 128, as it improved the convergence and generalization capabilities of the network \cite{keskar_2016}. The ADAM optimizer \cite{adam} is used with a decaying learning rate $\lambda_k = \dfrac{\lambda_0}{1 + \alpha k}$, where $\lambda_0 = 10^{-3}$, the decay $\alpha=10^{-3}$, and $k$ is the current iteration number. At each iteration during training, the network loss is computed as
\begin{equation}
    \mathcal{L} = \frac{1}{K}\sum_{k=1}^K\frac{1}{N}\sum_{i=1}^N \left\lVert \hat{\tilde{\boldsymbol{\eta}}}_i^{(k)} - \tilde{\boldsymbol{\eta}}_i^{(k)}\right\rVert_2^2 \quad, \label{eq:loss}
\end{equation}
where $\hat{\tilde{\boldsymbol{\eta}}}_i^{(k)}$ and $\tilde{\boldsymbol{\eta}}_i^{(k)}$ are the \textit{predicted} and \textit{true} target vectors, respectively, at the $i$-th time step in the $k$-th sample, $N$ is the length of the sequence of feature vectors input to the network, and $K$ is the number of samples in the mini batch. The loss in Equation \ref{eq:loss} is different than the loss used in \cite{Amato_2021} because this work considers sequence-o-sequence mapping, so the loss can be averaged over all time steps. Note that, during training, the network is fed the entire feature-vector sequence from each trajectory. The optimizer computes the gradient of Equation \ref{eq:loss} at each iteration and uses it to update the network's trainable parameters. To prevent gradient divergence, the $L^2$ norm of the gradient of each parameter in the network is limited to 1. Figure \ref{fig:train_loss} shows the training and validation loss across a 500-epoch training period. Throughout training, the validation loss stays at the same order of magnitude as the training loss, indicating that the network is capable of generalizing well to the validation data set.
Note that the validation loss is consistently less than the training loss. This is due to the use of dropout layers. As described in Subsection \ref{subsec:dropout}, the dropout layer may randomly drop some paths within the network during training, but during validation the dropout layers do not alter the network structure. The fact that the validation loss is consistently below the training loss illustrates the benefit of using dropout layers. The accuracy of the network's predictions are averaged over a separate test data set of 1000 trajectories to determine how well the network is able to generalize. Here, the error is computed as
\begin{equation}
    \Delta\rho = 100 \cdot\left|\frac{\hat{\rho}(h_j) - \rho(h_j)}{\rho(h_j)}\right| \quad,
\end{equation}
where $\hat{\rho}(h_j)$ is the predicted density at the target altitude $h_j$, and $\rho(h_j)$ is the truth density (from MarsGRAM) at the target altitude $h_j$.

\begin{figure}[htbp]
\begin{subfigure}{0.49\textwidth}\centering
  \includegraphics[width=\textwidth]{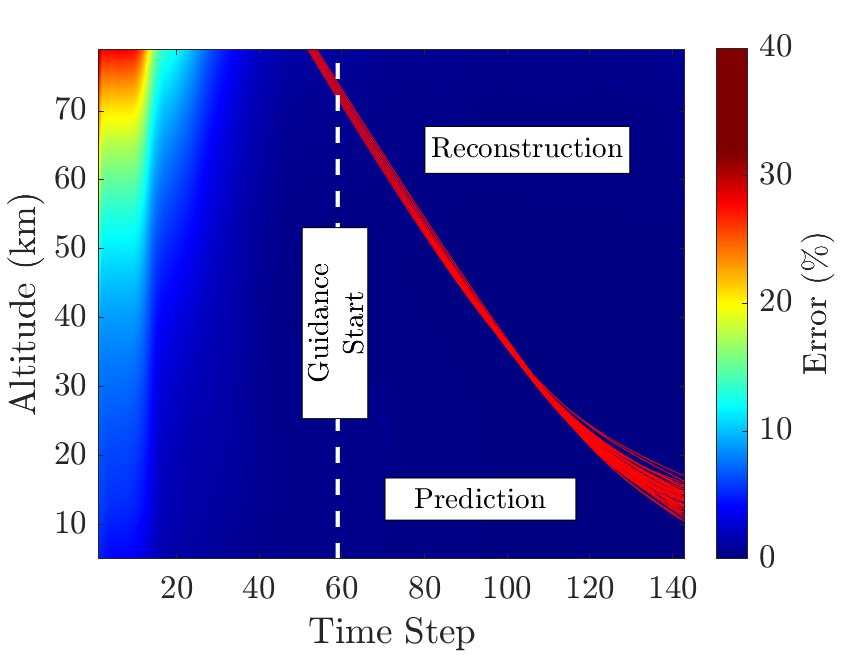}  
  \caption{Contour.}
  \label{fig:train_contour}
\end{subfigure}
\hfill
\begin{subfigure}{0.49\textwidth}\centering
  \includegraphics[width=\textwidth]{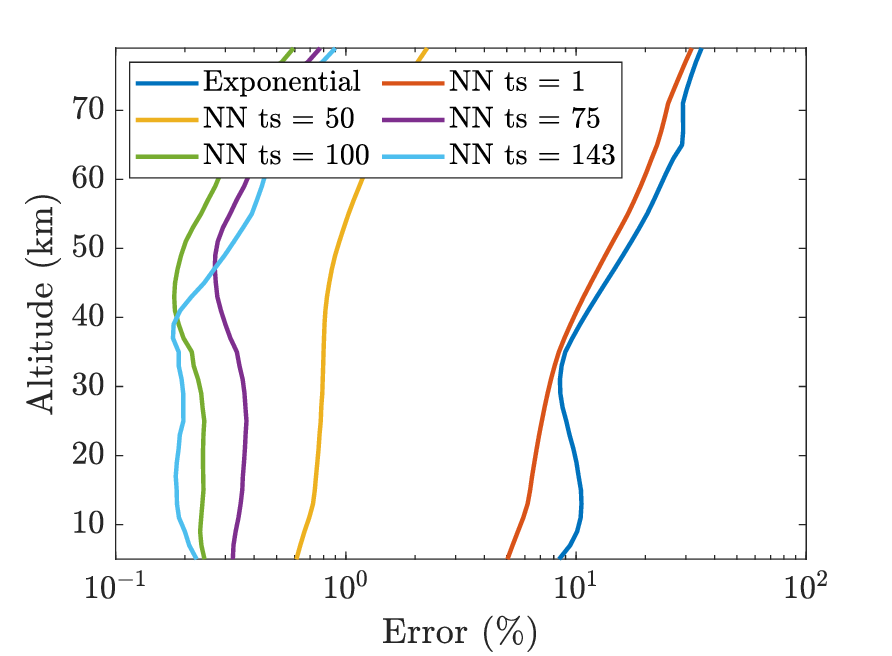}  
  \caption{Slices.}
  \label{fig:train_slice}
\end{subfigure}
\caption{Contour plot of error in density estimation (left) and slices at select time steps (right) averaged across test data set generated using an exponential density model in FNPEG. The overlaid red lines (left) show the altitude time history of 50 random trajectories within the validation data set.}
\label{fig:train_contour_slice}
\end{figure}

As the vehicles progresses along its trajectory, the vector of measurements passed to the LSTM grows in size. Recall that only measurements taken up to the current time step are passed to the LSTM to make density-profile predictions, so, as the vehicle collects more measurements, the LSTM should be able to produce more accurate density-profile predictions. To emulate the process of continuously receiving measurements during an entry trajectory, the network is fed feature-vector sequences that increase in length, and the predicted density profile is compared to the true profile output from MarsGRAM. The average error in density prediction across all the test cases is shown in Figure \ref{fig:train_contour_slice}, and 50 random trajectories from the test data set are overlaid for reference. Recall that the FNPEG algorithm is only used if the density is high enough, as indicated by the vertical white line in Figure \ref{fig:train_contour}.

As the length of the sequence in the validation data set input to the network increases, the network produces increasingly accurate estimations of the density. On average, by the time that the guidance algorithm is initiated, the error in the predicted density is $\leq 1\%$ for all altitudes $h_j$. Note that the error in density estimation is low at altitudes both \textit{below} and \textit{above} the reference trajectories overlaid in Figure \ref{fig:train_contour}, demonstrating that the network is able to both \textit{predict} the density through which it is going to fly and \textit{reconstruct} the density profile through which it has already flown. This is a significant improvement upon the work in \cite{Amato_2021}, as the network is able to accurately predict the density at all time steps in the trajectory, rather than just at the last. The loss in Figure \ref{fig:train_loss} could likely have been further decreased by training the network longer, but the accuracy in the network's predictions shown in Figure \ref{fig:train_contour_slice} is deemed acceptable.

\subsection{Curriculum Learning with FNPEG Guidance} \label{subsec:LSTM_FNPEG}
Recall that the data set used in training is generated by propagating a large number of trajectories guided by the FNPEG algorithm utilizing an \textit{exponential} density model. Because of this, the sequences of feature vectors used as training data are consistent with dynamics produced from an exponential model within the guidance system, not an LSTM-based model within the guidance system. Therefore, when the LSTM is implemented within FNPEG, the sequences of feature vectors will no longer originate from the same process as those generated using the exponential model, and the neural network's predictions will be significantly degraded compared to Figure \ref{fig:train_contour_slice}.

To achieve reliable density predictions when the LSTM is integrated within the FNPEG guidance loop, we train the network with an iterative \emph{curriculum learning} procedure \cite{wangSurveyCurriculumLearning2021}, summarized in Algorithm \ref{alg:training}. At the first iteration of the procedure, the LSTM network is trained on a data set generated with FNPEG using an exponential model, as shown in the remainder of this section. The accuracy of the network's predictions is quantified by computing the mean and standard deviation of the terminal range-to-go $|s_f|$ within a 5000-case Monte Carlo simulation in which the current LSTM network provides density profiles to the FNPEG prediction step. Terminal range-to-go is an effective measure of the network's predictive capabilities because it should be small if the guidance algorithm is supplied an accurate density model. The training and validation data set is then regenerated, as discussed in \ref{subsec:data_set_gen} and \ref{subsec:data_set_prep}, using the current LSTM model, and the network is retrained for 500 epochs. This process is repeated until the change in the terminal range-to-go mean $\Delta\mu(|s_f|)$ and standard deviation $\Delta\sigma(|s_f|)$ are both below a small tolerance $\epsilon=3\%$.
\RestyleAlgo{boxruled}
\LinesNumbered
\begin{algorithm}
\caption{Iterative LSTM Training Architecture}\label{alg:training}
Generate training data using exponential model in FNPEG\\
\While{done = false}{
\For(\tcp*[f]{Train network}){$k = 1:N_{\rm{epoch}}$}{ 
    $\mathcal{L} = \rm{loss(network)}$ \Comment*[r]{Compute loss}
    $\nabla\mathcal{L}$ = gradient($\mathcal{L}$) \Comment*[r]{Compute loss gradient}
    network = updateNet($\nabla\mathcal{L}$) \Comment*[r]{Update parameters}
}
Run Monte Carlo using trained network in FNPEG\\
Compute range-to-go stats\\
\eIf{$\Delta\mu(|s_f|) \leq \epsilon $ \rm{AND} $\Delta\sigma(|s_f|)\leq \epsilon$}{
    $done = true$ \Comment*[r]{Check if converged}
}{
    Regenerate training data using trained LSTM in FNPEG
}
}
\end{algorithm}

\begin{figure}[htbp]
\begin{subfigure}{0.49\textwidth}\centering
  \includegraphics[width=\textwidth]{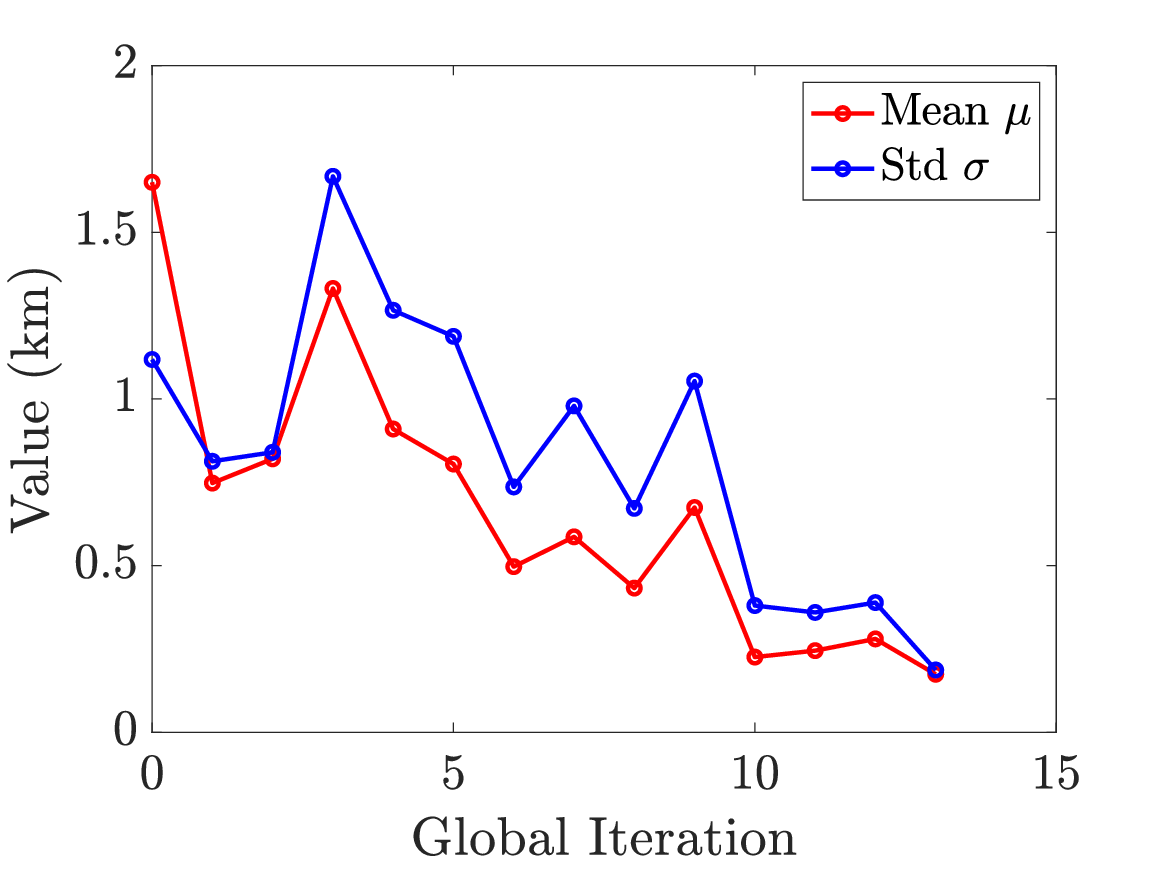}  
  \caption{Density profiles.}
  \label{fig:iterative_stats}
\end{subfigure}
\hfill
\begin{subfigure}{0.49\textwidth}\centering
  \includegraphics[width=\textwidth]{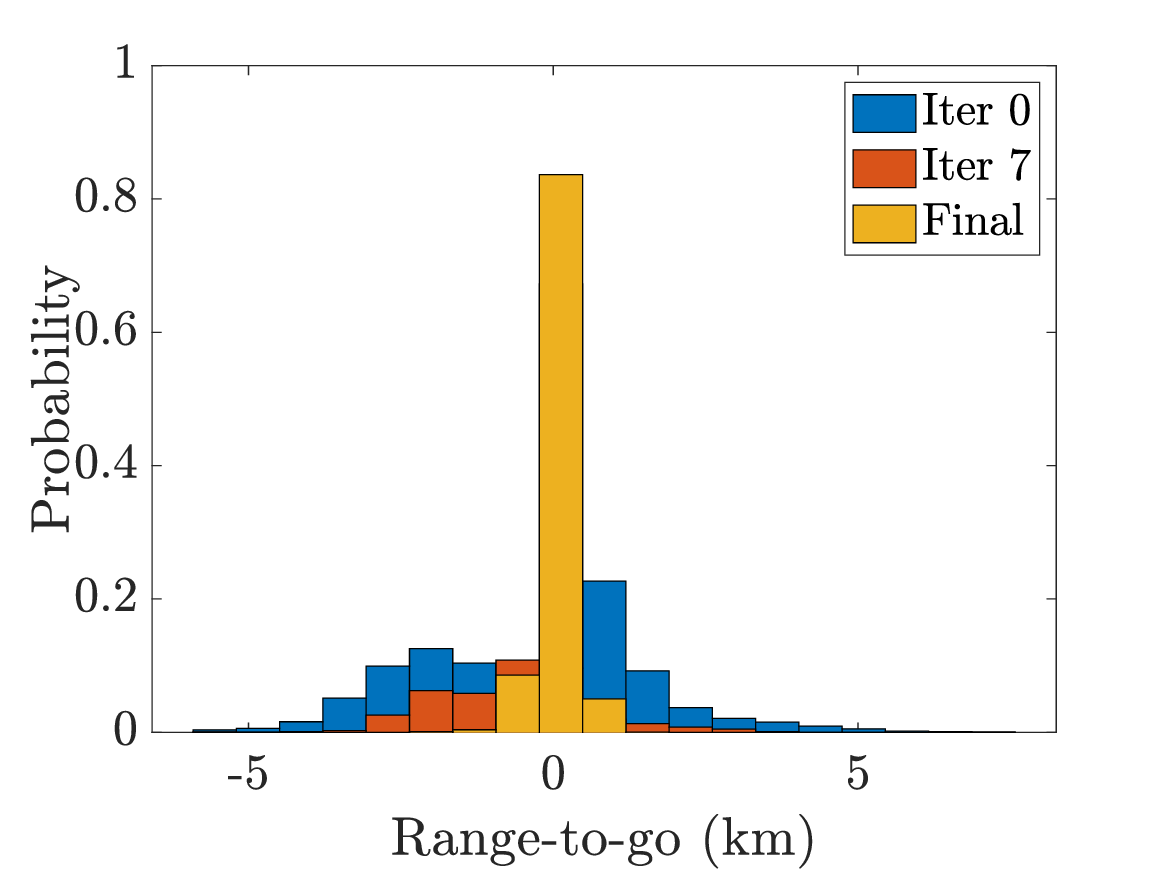}  
  \caption{Deviation from mean.}
  \label{fig:iterative_hist}
\end{subfigure}
\caption{Evolution of range-to-go statistics (left) and total distribution (right) during iterative LSTM training.}
\label{fig:iterative}
\end{figure}

Figure \ref{fig:iterative_stats} shows the evolution of the mean and standard deviation of the terminal range-to-go magnitude during the curriculum learning process and Figure \ref{fig:iterative_hist} shows a histogram of the signed terminal range-to-go at specific iterations. Here, a positive range-to-go represents an undershoot case, and a negative range-to-go represents an overshoot case.

Initially, the distribution in signed range-to-go is broad, as the baseline network was trained on feature sequences produced from an exponential density model within FNPEG. As the curriculum learning proceeds, both the mean and standard deviation of the terminal range-to-go decrease and the distribution in range-to-go becomes more confined and centered around zero. The increase in accuracy is explained by the training data containing more dynamically consistent feature-vector sequences from one outer-loop iteration to the next, as expected. Beyond thirteen iterations in the outer loop, the statistics in the terminal range-to-go do not change significantly, so the training process is terminated. At the end of the iterative training process, the mean and standard deviation of the terminal range-to-go magnitude are $\mu = 0.17$ km and $\sigma = 0.18$ km, respectively. 

\section{Comparison to State-of-the-Art Density Estimation Methods}\label{sec:results}
The performance of the FNPEG algorithm, defined by the terminal range-to-go, is assessed using three different density estimations techniques. In the first, FNPEG uses a single exponential density model (Equation \ref{eq:exp_dens}) during every guidance call. In the second, FNPEG receives the predictions at each guidance call from the trained LSTM, as done in subsection \ref{subsec:LSTM_FNPEG}, and in the third, FNPEG uses the exponential model that is augmented by a first-order fading-memory filter. Previous work has shown that the fading-memory filter is an effective approach to capturing a constant bias in density estimation \cite{Brunner_2008, Lu_2015}. The filter is used to estimate the ratios
\begin{equation}
    \rho_L = L/L^*,\qquad \rho_D = D/D^* \label{eq:fading_mem_ratio} \quad,
\end{equation}
where $L$ and $D$ are the sensed aerodynamic lift and drag accelerations, respectively, and $L^*$ and $D^*$ are the expected aerodynamic lift and drag accelerations, respectively. At the $n$-th guidance call, the ratios in Equation \ref{eq:fading_mem_ratio} are updated according to
\begin{eqnarray}
    \tilde{\rho}_L^{(n)} &=& \tilde{\rho}_L^{(n-1)} + \left(1-\beta\right)\left(\rho_L - \tilde{\rho}_L^{(n-1)}\right), \qquad 0<\beta<1 \\
    \tilde{\rho}_D^{(n)} &=& \tilde{\rho}_D^{(n-1)} + \left(1-\beta\right)\left(\rho_D - \tilde{\rho}_D^{(n-1)}\right), \qquad 0<\beta<1 \quad,
\end{eqnarray}
where $\rho_L$ and $\rho_D$ are computed using the current sensed acceleration, $\tilde{\rho}_D^{(n-1)}$ and $\tilde{\rho}_D^{(n-1)}$ are the ratios taken from the previous guidance call, and $\beta$ is a tuning parameter. For all simulations, a constant value of $\beta=0.9$ is assumed. The filters are initialized by setting $\tilde{\rho}_L^{(0)}=\tilde{\rho}_D^{(0)}=1.0$. Then, the values of $\tilde{\rho}_L^{(n)}$ and $\tilde{\rho}_D^{(n)}$ are used to scale the expected lift and drag accelerations in Equation \ref{eq:lon_dyn} during the prediction step of FNPEG.

\begin{figure}[htbp]
    \centering
    \includegraphics[width=0.7\textwidth]{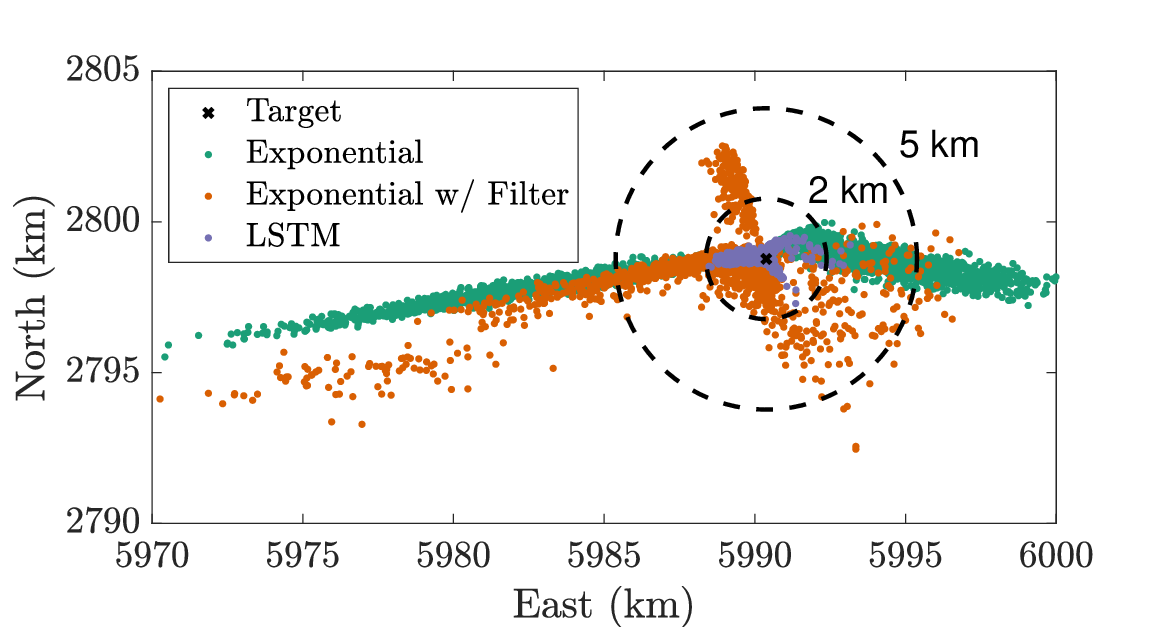}
    \caption{Terminal accuracy results from 5000-case Monte Carlo simulations using each density model in FNPEG.}
    \label{fig:term_ellipse}
\end{figure}

\begin{figure}[htbp]
    \centering
    \hspace{-0.25 in}\includegraphics[width=0.6\textwidth]{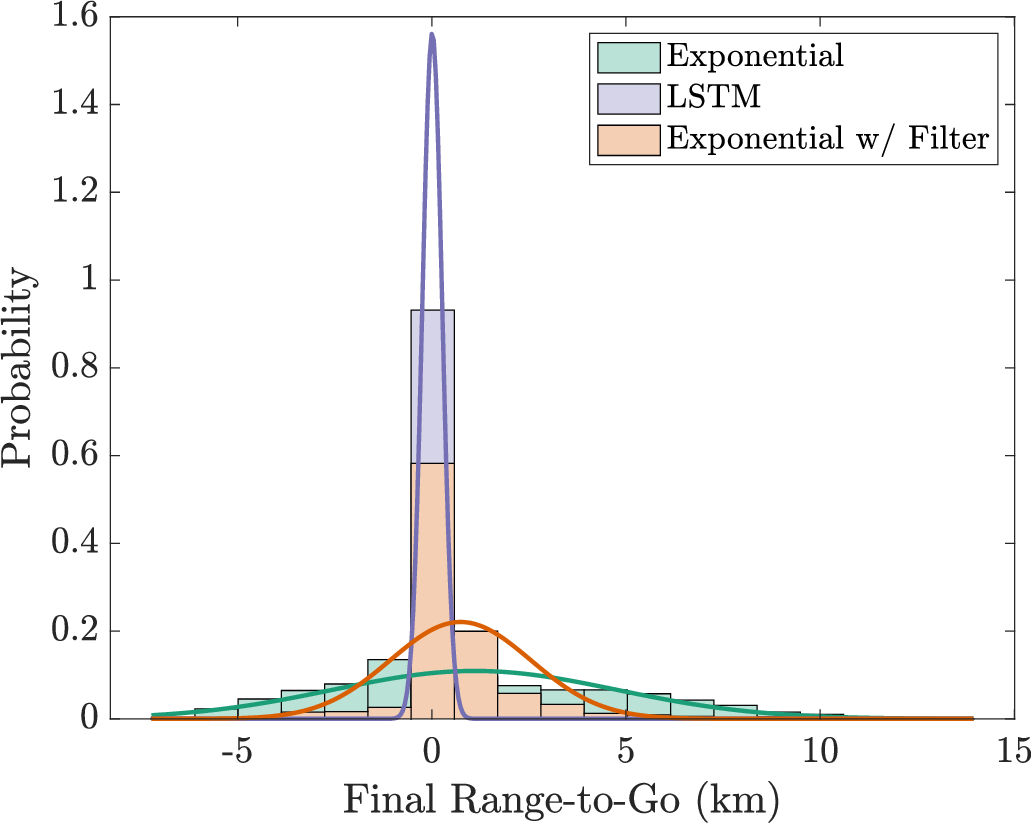}
    \caption{Histogram of final range-to-go from 5000-case Monte Carlo simulations using each density model in FNPEG. Solid lines represent Gaussian probability density functions fit to each data set.}
    \label{fig:s_hist}
\end{figure}

For each density estimation technique, a 5000-case Monte Carlo analysis is used to assess its impact on the guidance system performance. Similar to the data set generation in \ref{subsec:data_set_gen}, the truth atmospheric density profile is varied by drawing random samples for the dust level, mean wave offset, and GRAM random seed from the distributions in Table \ref{tab:gram}. Each trajectory is integrated until the target final energy is reached. Figure \ref{fig:term_ellipse} shows the distribution of final positions for each density estimation method, and Figure \ref{fig:s_hist} shows the corresponding histogram of signed range-to-go. The statistics of the range-to-go magnitude are summarized in Table \ref{tab:s_stats}. The solid lines in Figure \ref{fig:s_hist} correspond to Gaussian probability density functions fit to each histogram distribution.

Clearly, the LSTM-augmented FNPEG significantly outperforms the other two models. The overshoot and undershoot cases depicted in Figure \ref{fig:term_ellipse} are explained by each model either under- or overpredicting, respectively, the true density computed from GRAM. For all 5000 cases, the LSTM-based model produces terminal range-to-go values that are completely contained by a 2-km radius circle, while the other two methods struggle to do so. The method using the first-order fading-memory filter does a much better job compared to the exponential model, but it is still unable to reduce its final distribution in range-to-go as effectively as the LSTM model. This is likely due to the assumption that the filter is designed to account for constant biases in density, and Figure \ref{fig:gram_dev} demonstrates that this assumption is not reflected in reality. The large population of over- and undershoot cases using the exponential and fading-memory filter methods is dominated by the failure of the exponential model to capture day-of-flight variations in density (Figure \ref{fig:gram_dens}), particularly at low altitudes.

\begin{table}[htbp]
    \centering
    \caption{Statistics for terminal range-to-go magnitude for each density estimation method.}
    \begin{tabular}{l c c c c}
    \toprule\toprule
        Method & Mean (km) & $\sigma$ (km) & 1$^{\rm{st}}$-percentile (km) & 99$^{\rm{th}}$-percentile (km)\\ \midrule
         Exponential & 2.93 & 2.44 & 0.04 & 10.13\\
         Exponential w/ Filter & 1.02 & 1.66 & 0.02 & 8.03\\
         LSTM & 0.17  & 0.18 & 0.01 & 1.23\\
         \bottomrule\bottomrule
    \end{tabular}
    \label{tab:s_stats}
\end{table}

\FloatBarrier
\subsection{Noise in Measurements}\label{subsec:noise}
The previous analysis assumed that the measurements used to train the LSTM, and the measurements passed to the LSTM at each guidance cycle, were noiseless. Clearly, this is not representative of the true measurements received by an actual guidance system. Therefore, to assess the ability of the LSTM to adapt to noisy measurements, random i.i.d.~noise is introduced into the feature-vector sequences. For the six spherical state and three Cartesian acceleration components of each feature vector, the noise level was modeled according to the recommendations in Reference \cite{dwyer2019defining}, and for the pressure-measurement feature, the noise level was modeled according to the values in  Reference \cite{Hwang_2016}. Table \ref{tab:noise} summarizes the noise level for each component of the feature vector.
\begin{table}[htbp]
    \centering
    \caption{Noise levels for feature vector.}
    \begin{tabular}{l c c c c c c c c}
    \toprule \toprule
         Feature   & $r$ (m) & $\theta$ (deg) & $\phi$ (deg) & $v$ (m/s) & $\gamma$ (deg) & $\psi$ (deg) &  $a_i$ (g) & $P_{0,2}$ (\%)  \\ \midrule
         $3\sigma$ & 5.0 & $8.4\cdot10^{-5}$ & $8.4\cdot10^{-5}$ & 1.0 & 0.01 & 0.01 & $10^{-7}$ & 1.0 \\ \bottomrule\bottomrule
    \end{tabular}
    \label{tab:noise}
\end{table}

\begin{figure}[htbp]
    \centering
    \includegraphics[width=0.75\textwidth]{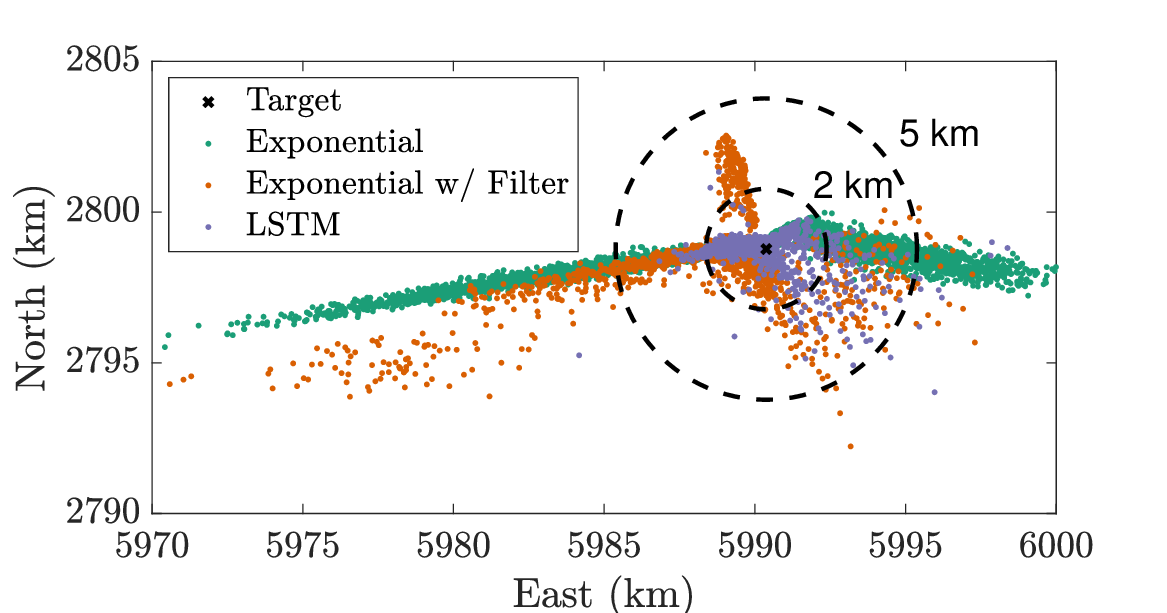}
    \caption{Terminal accuracy results from 5000-case Monte Carlo simulations using each density model in FNPEG with noisy measurements.}
    \label{fig:term_ellipse_noisy}
\end{figure}

\begin{figure}[htbp]
    \centering
    \hspace{-0.25 in}\includegraphics[width=0.6\textwidth]{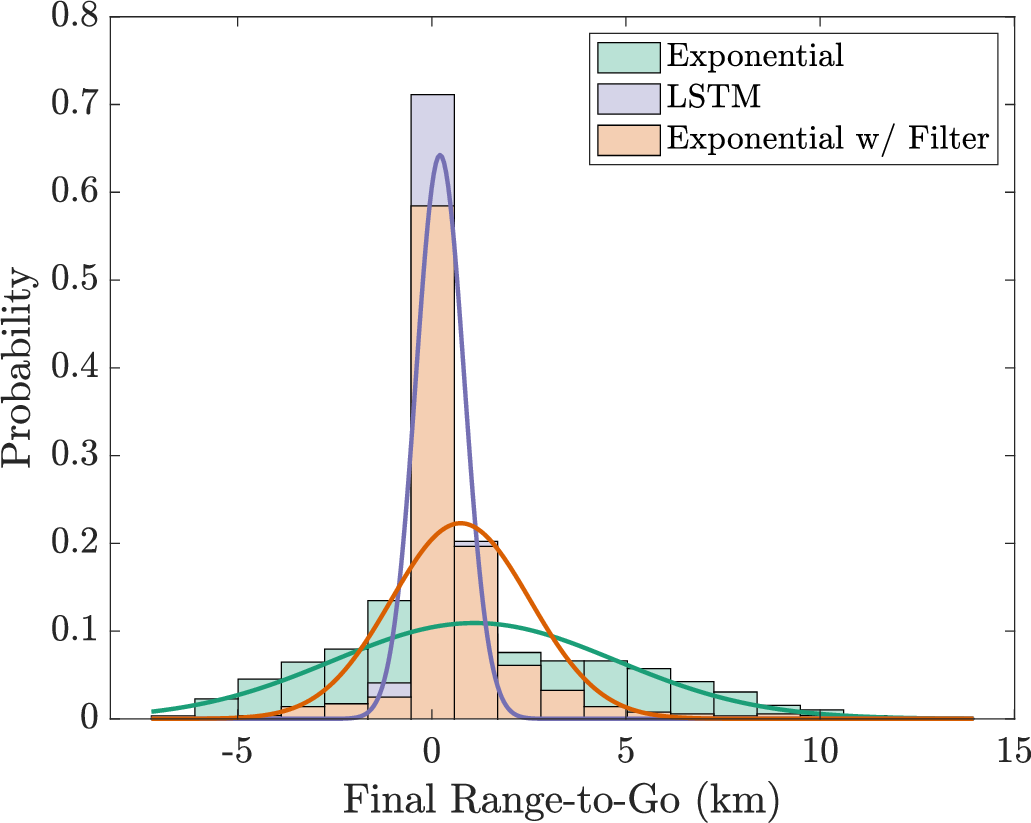}
    \caption{Histogram of signed range-to-go from 5000-case Monte Carlo simulations using each density model in FNPEG with noisy measurements. Solid lines represent Gaussian probability density functions fit to each data set.}
    \label{fig:s_hist_noisy}
\end{figure}

At each guidance cycle, zero-mean normal distributions with variance levels defined by Table \ref{tab:noise} are sampled to perturb the nominal measurements comprising the current feature vector in the feature-vector sequence. The previously trained LSTM used to produce the results in Figures \ref{fig:term_ellipse} and \ref{fig:s_hist} is retrained with the noisy measurements for a single 500-epoch training period. The introduction of noise into the feature vectors essentially adds a regularization term into the network training, which should improve the network's robustness in the presence of noisy data. Similar to the analysis with noiseless measurements, a 5000-case Monte Carlo analysis was run for both the LSTM density-estimation method and the first-order fading-memory density estimation method to assess the impact of noisy measurements on the guidance system's performance. Figure \ref{fig:term_ellipse_noisy} shows the distribution of final positions for each density estimation method with noisy measurements, and Figure \ref{fig:s_hist_noisy} shows the corresponding histogram of signed range-to-go. The statistics of the range-to-go magnitude using noisy measurements are summarized in Table \ref{tab:s_stats_noisy}. The solid lines in Figure \ref{fig:s_hist_noisy} correspond to Gaussian probability density functions fit to each histogram distribution.

When compared to the noiseless case, both the LSTM and the first-order filter exhibit slightly less accurate terminal range-to-go distributions. However, the LSTM method with noisy data still significantly outperforms both the noisy \textit{and} noiseless first-order filter method. The slightly larger distribution for the terminal range-to-go using the LSTM is expected, as the regularization of the network through the introduction of noise should produce a network that is slightly less accurate, but has greater generalization capabilities, than the network trained on deterministic measurements. 
\begin{table}[htbp]
    \centering
    \caption{Statistics for terminal range-to-go magnitude for each density estimation method using noisy measurements.}
    \begin{tabular}{l c c c c}
    \toprule\toprule
        Method & Mean (km) & $\sigma$ (km) & 1$^{\rm{st}}$-percentile (km) & 99$^{\rm{th}}$-percentile (km)\\ \midrule
         Exponential & 2.93 & 2.44 & 0.04 & 10.13\\
         Exponential w/ Filter & 1.03 & 1.66 & 0.02 & 9.15\\
         LSTM & 0.44  & 0.48 & 0.01 & 2.46\\
         \bottomrule\bottomrule
    \end{tabular}
    \label{tab:s_stats_noisy}
\end{table}
\section{Conclusions}\label{sec:conclusions}
This work presented a deep learning approach to estimate atmospheric density profiles for use in planetary entry guidance problems. A long short-term memory (LSTM) neural network was trained to learn the mapping between measurements available onboard an entry vehicle and the density profile through which it is flying. Specifically, the network utilizes the vehicle's spherical state, Cartesian acceleration components and surface-pressure measurements to predict the density as a function of altitude.

Initially, a 5000-case Monte Carlo analysis of a Martian entry mission, in which the FNPEG algorithm utilized an exponential density model and the true density profiles were sampled from MarsGRAM, was used to generate the network training data. The network was trained to minimize the difference, quantified by using an $L^2$-norm, between the network predictions and the corresponding density profiles from MarsGRAM. The trained network demonstrated excellent predictive capabilities, with average errors < 1\%, for almost all time steps in the trajectory. Additionally, the network was shown to be capable of \textit{predicting} future density profiles and \textit{reconstructing} density profiles through which the vehicle had already flown.

To integrate the network within the FNPEG guidance algorithm, a curriculum learning procedure was devised. During each iteration of the procedure, the training data set was regenerated using the LSTM trained from the previous iteration. The recursive procedure produced significant refinement in the targeting accuracy of the FNPEG algorithm, with the final mean and standard deviation in the terminal range-to-go magnitude being $\mu=0.17 \,\text{km}$ and $\sigma=0.18\,\text{km}$, respectively. When compared to the performance of the FNPEG algorithm using a first-order fading-memory filter to estimate the density, the LSTM model produced an order of magnitude improvement in terminal range-to-go accuracy. The increase in terminal accuracy of FNPEG using the LSTM was even larger when compared to the accuracy of FNPEG using an exponential density model. The significant improvement in guidance performance was maintained, even when noise was introduced to the onboard measurements.  As such, the three primary contributions and conclusions from this work were:
\begin{enumerate}
    \item An LSTM-based model was shown to produce accurate density predictions at all time steps along an entry trajectory.
    \item The LSTM network was successfully integrated into the FNPEG guidance algorithm through a curriculum learning approach, significantly increasing its performance.
    \item The LSTM network was capable of producing accurate density predictions in the presence of noisy measurements.
\end{enumerate}

Future work will investigate improvements to the entry guidance algorithm itself, extend the network's application to missions involving skip entry and aerocapture, and consider the quantification of uncertainty on atmospheric density. 

\section*{Acknowledgements}
This work was funded by NASA Space Technology Graduate Research Opportunity Fellowship award 80NSSC22K1186.

\bibliography{references}

\begin{thebibliography}{36}
\newcommand{\enquote}[1]{``#1''}
\providecommand{\natexlab}[1]{#1}
\providecommand{\url}[1]{\texttt{#1}}
\providecommand{\urlprefix}{URL }
\expandafter\ifx\csname urlstyle\endcsname\relax
  \providecommand{\doi}[1]{\discretionary{}{}{}https://doi.org/#1}\else
  \providecommand{\doi}[1]{\discretionary{}{}{}\urlstyle{rm}\url{https://doi.org/#1}}\fi

\bibitem[{Engelund et~al.(2008)Engelund, Powell, and Tolson}]{engelund2008atmospheric}
Engelund, W., Powell, R., and Tolson, R., \enquote{Atmospheric Modeling Challenges and Measurement Requirements for Mars Entry, Descent and Landing,} \emph{LPI Contributions}, Vol. 1447, 2008, p. 9025.

\bibitem[{Way et~al.(2022)Way, Dutta, Zumwalt, and Blette}]{way2022assessment}
Way, D.~W., Dutta, S., Zumwalt, C., and Blette, D., \enquote{Assessment of the Mars 2020 Entry, Descent, and Landing Simulation,} \emph{AIAA SciTech 2022 Forum}, 2022, p. 0421.

\bibitem[{Dutta et~al.(2023)Dutta, Karlgaard, Kass, Mischna, and Villar~III}]{dutta2023postflight}
Dutta, S., Karlgaard, C.~D., Kass, D., Mischna, M., and Villar~III, G.~G., \enquote{Postflight Analysis of Atmospheric Properties from Mars 2020 Entry, Descent, and Landing,} \emph{Journal of Spacecraft and Rockets}, Vol.~60, No.~3, 2023, pp. 1022--1033.

\bibitem[{Prakash et~al.(2008)Prakash, Burkhart, Chen, Comeaux, Guernsey, Kipp, Lorenzoni, Mendeck, Powell, Rivellini et~al.}]{prakash2008mars}
Prakash, R., Burkhart, P.~D., Chen, A., Comeaux, K.~A., Guernsey, C.~S., Kipp, D.~M., Lorenzoni, L.~V., Mendeck, G.~F., Powell, R.~W., Rivellini, T.~P., et~al., \enquote{Mars Science Laboratory entry, descent, and landing system overview,} \emph{2008 IEEE Aerospace Conference}, IEEE, 2008, pp. 1--18.

\bibitem[{Nelessen et~al.(2019)Nelessen, Sackier, Clark, Brugarolas, Villar, Chen, Stehura, Otero, Stilley, Way et~al.}]{nelessen2019mars}
Nelessen, A., Sackier, C., Clark, I., Brugarolas, P., Villar, G., Chen, A., Stehura, A., Otero, R., Stilley, E., Way, D., et~al., \enquote{Mars 2020 entry, descent, and landing system overview,} \emph{2019 IEEE Aerospace Conference}, IEEE, 2019, pp. 1--20.

\bibitem[{Lugo et~al.(2020)Lugo, Powell, and Dwyer-Cianciolo}]{lugo2020overview}
Lugo, R.~A., Powell, R., and Dwyer-Cianciolo, A.~M., \enquote{Overview of a generalized numerical predictor-corrector targeting guidance with application to human-scale mars entry, descent, and landing,} \emph{AIAA Scitech 2020 Forum}, 2020, p. 0846.

\bibitem[{Vinh et~al.(1980)Vinh, Buseman, and Culp}]{Vinh}
Vinh, N.~X., Buseman, A., and Culp, R.~D., \emph{Hypersonic and Planetary Entry Flight Mechanics}, The University of Michigan Press, 1980.

\bibitem[{Perot and Rousseau(2002)}]{perot2002importance}
Perot, E., and Rousseau, S., \enquote{Importance of an on-board estimation of the density scale height for various aerocapture guidance algorithms,} \emph{AIAA/AAS Astrodynamics Specialist Conference and Exhibit}, 2002, p. 4734.

\bibitem[{Masciarelli et~al.(2000)Masciarelli, Rousseau, Fraysse, and Perot}]{masciarelli2000analytic}
Masciarelli, J., Rousseau, S., Fraysse, H., and Perot, E., \enquote{An analytic aerocapture guidance algorithm for the Mars Sample Return Orbiter,} \emph{Atmospheric flight mechanics conference}, 2000, p. 4116.

\bibitem[{Brunner and Lu(2008)}]{Brunner_2008}
Brunner, C.~W., and Lu, P., \enquote{Skip entry trajectory planning and guidance,} \emph{Journal of Guidance, Control, and Dynamics}, Vol.~31, No.~5, 2008, pp. 1210--1219.

\bibitem[{Lu et~al.(2015)Lu, Cerimele, Tigges, and Matz}]{Lu_2015}
Lu, P., Cerimele, C.~J., Tigges, M.~A., and Matz, D.~A., \enquote{Optimal aerocapture guidance,} \emph{Journal of Guidance, Control, and Dynamics}, Vol.~38, No.~4, 2015, pp. 553--565.

\bibitem[{Dutta et~al.(2014)Dutta, Braun, and Karlgaard}]{dutta2014uncertainty}
Dutta, S., Braun, R.~D., and Karlgaard, C.~D., \enquote{Uncertainty quantification for mars entry, descent, and landing reconstruction using adaptive filtering,} \emph{Journal of Spacecraft and Rockets}, Vol.~51, No.~3, 2014, pp. 967--977.

\bibitem[{Heidrich et~al.(2022)Heidrich, Holzinger, and Braun}]{heidrich2022optimal}
Heidrich, C.~R., Holzinger, M.~J., and Braun, R.~D., \enquote{Optimal information filtering for robust aerocapture trajectory generation and guidance,} \emph{Journal of Spacecraft and Rockets}, Vol.~59, No.~2, 2022, pp. 524--537.

\bibitem[{Roelke et~al.(2023)Roelke, McMahon, Braun, and Hattis}]{roelke2023atmospheric}
Roelke, E., McMahon, J.~W., Braun, R.~D., and Hattis, P.~D., \enquote{Atmospheric Density Estimation Techniques for Aerocapture,} \emph{Journal of Spacecraft and Rockets}, 2023, pp. 1--15.

\bibitem[{Tracy et~al.(2023)Tracy, Falcone, and Manchester}]{tracy2023robust}
Tracy, K.~S., Falcone, G., and Manchester, Z., \enquote{Robust Entry Guidance with Atmospheric Adaptation,} \emph{AIAA SCITECH 2023 Forum}, 2023, p. 0301.

\bibitem[{Arridge et~al.(2019)Arridge, Maass, {\"O}ktem, and Sch{\"o}nlieb}]{arridge2019solving}
Arridge, S., Maass, P., {\"O}ktem, O., and Sch{\"o}nlieb, C.-B., \enquote{Solving inverse problems using data-driven models,} \emph{Acta Numerica}, Vol.~28, 2019, pp. 1--174.

\bibitem[{Wagner et~al.(2011)Wagner, Wilhite, Stanley, and Powell}]{wagner2011adaptive}
Wagner, J., Wilhite, A., Stanley, D., and Powell, R., \enquote{An adaptive real time atmospheric prediction algorithm for entry vehicles,} \emph{3rd AIAA Atmospheric Space Environments Conference}, 2011, p. 3200.

\bibitem[{Amato and McMahon(2021)}]{Amato_2021}
Amato, D., and McMahon, J.~W., \enquote{Deep Learning Method for Martian Atmosphere Reconstruction,} \emph{Journal of Aerospace Information Systems}, Vol.~18, No.~10, 2021, pp. 728--738.
\newblock \doi{10.2514/1.I010922}.

\bibitem[{Goodfellow et~al.(2016)Goodfellow, Bengio, and Courville}]{Goodfellow-et-al-2016}
Goodfellow, I., Bengio, Y., and Courville, A., \emph{Deep Learning}, MIT Press, 2016.
\newblock \url{http://www.deeplearningbook.org}.

\bibitem[{Hochreiter(1998)}]{Hochreiter_1998}
Hochreiter, S., \enquote{Recurrent neural net learning and vanishing gradient,} \emph{International Journal Of Uncertainity, Fuzziness and Knowledge-Based Systems}, Vol.~6, No.~2, 1998, pp. 107--116.

\bibitem[{Hochreiter and Schmidhuber(1997)}]{Hochreiter_1997}
Hochreiter, S., and Schmidhuber, J., \enquote{{Long Short-Term Memory},} \emph{Neural Computation}, Vol.~9, No.~8, 1997, pp. 1735--1780.
\newblock \doi{10.1162/neco.1997.9.8.1735}.

\bibitem[{Sherstinsky(2020)}]{Sherstinsky_2020}
Sherstinsky, A., \enquote{Fundamentals of Recurrent Neural Network ({RNN}) and Long Short-Term Memory ({LSTM}) network,} \emph{Physica D: Nonlinear Phenomena}, Vol. 404, 2020, p. 132306.
\newblock \doi{10.1016/j.physd.2019.132306}.

\bibitem[{Sutskever et~al.(2014)Sutskever, Vinyals, and Le}]{Sutskever_2014}
Sutskever, I., Vinyals, O., and Le, Q.~V., \enquote{Sequence to sequence learning with neural networks,} \emph{Advances in neural information processing systems}, Vol.~27, 2014.

\bibitem[{Srivastava et~al.(2014)Srivastava, Hinton, Krizhevsky, Sutskever, and Salakhutdinov}]{Srivastava_2014}
Srivastava, N., Hinton, G., Krizhevsky, A., Sutskever, I., and Salakhutdinov, R., \enquote{Dropout: A Simple Way to Prevent Neural Networks from Overfitting,} \emph{Journal of Machine Learning Research}, Vol.~15, No.~56, 2014, pp. 1929--1958.
\newblock \urlprefix\url{http://jmlr.org/papers/v15/srivastava14a.html}.

\bibitem[{Justh et~al.(2021)Justh, Cianciolo, and Hoffman}]{mars_gram2021}
Justh, H.~L., Cianciolo, A. M.~D., and Hoffman, J., \enquote{Mars Global Reference Atmospheric Model (Mars-GRAM): User Guide,} Tech. rep., National Aeronautics and Space Administration, NASA/TM-20210023957, November 2021.

\bibitem[{Hughes et~al.(2011)Hughes, Cheatwood, Dillman, Calomino, Wright, DelCorso, and Calomino}]{hughes2011hypersonic}
Hughes, S., Cheatwood, F., Dillman, R., Calomino, A., Wright, H., DelCorso, J., and Calomino, A., \enquote{Hypersonic inflatable aerodynamic decelerator (HIAD) technology development overview,} \emph{21st AIAA Aerodynamic Decelerator Systems Technology Conference and Seminar}, 2011, p. 2524.

\bibitem[{Anderson(2019)}]{Anderson_2019_ch3}
Anderson, J.~D., \emph{Hypersonic and High-Temperature Gas Dynamics, Third Edition}, AIAA Education Series, 2019.
\newblock \doi{10.2514/5.9781624105142.0055.0106}.

\bibitem[{Lu(2014)}]{Lu_2014}
Lu, P., \enquote{Entry Guidance: A Unified Method,} \emph{Journal of Guidance, Control, and Dynamics}, Vol.~37, No.~3, 2014, pp. 713--728.
\newblock \doi{10.2514/1.62605}.

\bibitem[{Lu(2008)}]{Lu_2008}
Lu, P., \enquote{Predictor-corrector entry guidance for low-lifting vehicles,} \emph{Journal of Guidance, Control, and Dynamics}, Vol.~31, No.~4, 2008, pp. 1067--1075.

\bibitem[{Hwang et~al.(2016)Hwang, Bose, Wright, White, Schoenenberger, Santos, Karlgaard, Kuhl, Oishi, and Trombetta}]{Hwang_2016}
Hwang, H., Bose, D., Wright, H., White, T.~R., Schoenenberger, M., Santos, J., Karlgaard, C.~D., Kuhl, C., Oishi, T., and Trombetta, D., \enquote{Mars 2020 Entry, Descent, and Landing Instrumentation (MEDLI2),} 46th AIAA Thermophysics Conference, 2016.
\newblock \doi{10.2514/6.2016-3536}.

\bibitem[{White et~al.(2022)White, Mahzari, Miller, Tang, Karlgaard, Alpert, Wright, and Kuhl}]{White_2022}
White, T.~R., Mahzari, M., Miller, R.~A., Tang, C.~Y., Karlgaard, C.~D., Alpert, H., Wright, H.~S., and Kuhl, C., \enquote{Mars Entry Instrumentation Flight Data and Mars 2020 Entry Environments,} AIAA SCITECH 2022 Forum, 2022.
\newblock \doi{10.2514/6.2022-0011}.

\bibitem[{Montavon et~al.(2012)Montavon, Orr, and Mller}]{Montavon_2012}
Montavon, G., Orr, G., and Mller, K.-R., \emph{Neural Networks: Tricks of the Trade}, 2\textsuperscript{nd} ed., Springer Publishing Company, Incorporated, 2012.

\bibitem[{Keskar et~al.(2016)Keskar, Mudigere, Nocedal, Smelyanskiy, and Tang}]{keskar_2016}
Keskar, N.~S., Mudigere, D., Nocedal, J., Smelyanskiy, M., and Tang, P. T.~P., \enquote{On large-batch training for deep learning: Generalization gap and sharp minima,} \emph{arXiv preprint arXiv:1609.04836}, 2016.

\bibitem[{Kingma and Ba(2014)}]{adam}
Kingma, D.~P., and Ba, J., \enquote{Adam: A method for stochastic optimization,} \emph{arXiv preprint arXiv:1412.6980}, 2014.

\bibitem[{Wang et~al.(2021)Wang, Chen, and Zhu}]{wangSurveyCurriculumLearning2021}
Wang, X., Chen, Y., and Zhu, W., \enquote{A {{Survey}} on {{Curriculum Learning}},} \emph{IEEE Transactions on Pattern Analysis and Machine Intelligence}, 2021.
\newblock \doi{10.1109/TPAMI.2021.3069908}.

\bibitem[{Dwyer-Cianciolo et~al.(2019)Dwyer-Cianciolo, Karlgaard, Woffinden, Lugo, Tynis, Sostaric, Striepe, Powell, and Carson}]{dwyer2019defining}
Dwyer-Cianciolo, A.~M., Karlgaard, C.~D., Woffinden, D., Lugo, R.~A., Tynis, J., Sostaric, R.~R., Striepe, S., Powell, R., and Carson, J.~M., \enquote{Defining navigation requirements for future missions,} \emph{AIAA SciTech 2019 forum}, 2019, p. 0661.

\end{thebibliography}
\end{document}